\documentclass[authoryear]{article}       
\usepackage{graphicx}
\usepackage{amsmath,amssymb,bm,graphicx}
\usepackage{float}
\usepackage{color,grffile}
\usepackage{times,subcaption}
\usepackage{rotating}
\usepackage{mathtools}
\usepackage{hyperref}
\usepackage{bm}
\usepackage{natbib}
\usepackage[plain,noend]{algorithm2e}
\usepackage{subcaption}
\usepackage{lineno}
\usepackage{geometry}

\newcommand{\Nbsim}{{S}}
\newcommand{\Nbvar}{{p}}
\newcommand{\bfX}{{\bold{X}}}
\newcommand{\bfR}{{\bold{R}}}
\newcommand{\bfM}{{\bold{M}}}
\newcommand{\bfH}{{\bold{H}}}
\newcommand{\nbvar}{{\ell}}
\newcommand{\Nbgroup}{{K}}

\newcommand{\Nbpart}{{J}}
\newcommand{\nbpart}{{j}}
\newcommand{\Nbind}{{n}}
\newcommand{\nbind}{{i}}
\newcommand{\Nbboot}{{C}}
\newcommand{\nbboot}{{c}}
\newcommand{\Nbtab}{{M}}
\newcommand{\nbtab}{{m}}
\newcommand{\QI}{{Q}}

\begin{document}

\title{Clustering with missing data: which equivalent for Rubin's rules?}
\author{Vincent Audigier\footnote{vincent.audigier@cnam.fr}, Ndeye Niang \footnote{n-deye.niang\_ keita@cnam.fr} \\ {\small CNAM, Laboratoire CEDRIC-MSDMA, 2 rue Cont\'e, 75003 Paris, France}
}

\maketitle

\textbf{Abstract} 
Multiple imputation (MI) is a popular method for dealing with missing values. However, the suitable way for applying clustering after MI remains unclear: how to pool partitions? How to assess the clustering instability when data are incomplete? By answering both questions, this paper proposed a complete view of clustering with missing data using MI. The problem of partitions pooling is here addressed using consensus clustering while, based on the bootstrap theory, we explain how to assess the instability related to observed and missing data. The new rules for pooling partitions and instability assessment are theoretically argued and extensively studied by simulation. Partitions pooling improves accuracy, while measuring instability with missing data enlarges the data analysis possibilities: it allows assessment of the dependence of the clustering to the imputation model, as well as a convenient way for choosing the number of clusters when data are incomplete, as illustrated on a real data set.

\vspace{1cm}


\section{Introduction}
Clustering individuals is an essential task for data science. Clustering aims at partitioning a sample of individuals in several groups (clusters) so that individuals in a same cluster are similar from a multidimensional point of view, while individuals in separate clusters are different. $k-$means clustering \citep{Forgy65}, partitioning around medoids \citep{Kaufman90}, clustering by mixture models \citep{mclachlan88} or hierarchical clustering \citep{Ward63} are some popular methods for building this partition.

However, data are often incomplete and clustering algorithms cannot be directly applied on incomplete data. A common strategy to deal with missing values in data analysis consists in using multiple imputation \citep{Rubin76,Rubin87,Schafer97}. Multiple imputation (MI) consists of 3 steps: 1) the imputation of the data set according to an imputation model several times 2) the analysis of each imputed data set according to a substantive model 3) the pooling of the analysis results according to the Rubin's rules. Such methods are mainly used for inference in linear models (see \citet{Marshall09} for other uses), but not for clustering. For instance, for applying a regression model on incomplete quantitative data: 1) data can be imputed according to a Gaussian model \citep{Schafer97}, 2) the regression model is fit on each imputed data set, leading to several estimates of the regression coefficients and their associated variance, 3) regression estimates are averaged and the associated variance is computed. Thus, despite missing values, MI yields a unique estimate of substantive model parameters and an uncertainty measure, which is expressed as a variance estimate. One major issue for applying clustering after MI is how to apply an equivalent of Rubin's rules in this context, \textit{i.e.} how to apply step 3) ? \citet{Basagana13, Faucheux, Bruckers17} brought some answers in terms of partitions pooling, but the question of the uncertainty measure has not been discussed.

Uncertainty in clustering refers to \textit{(in)stability} and results in various Voronoi tessellations of the metric space. These variations can cover many aspects: the used clustering algorithm, its initialization, the chosen number of clusters, etc. Here, we focus on another aspect which is the stability related to sampling. Note that we distinguish it 
to the probability in the assignment of an individual to a cluster (easily obtained for model-based clustering methods), which assesses the uncertainty in cluster assignment, but would remain even if the sample were fixed. See \citet{Dudoit03} or \citet{Bruckers17} for related works. 

When data are complete, several resampling techniques have been proposed to assess the clustering stability \citep{Henning07}. One main advantage of these methods consists in being relevant for both distance-based and model-based clustering methods. They are generally motivated by the determination of the number of clusters. The rationale is that a ``too" large number of clusters should lead to a significant increase of instability.  \citet{Jain87, Wang10, Fang12, Mourer20} proposed several approaches in this line. In particular, \citet{Wang10} proposed a measure of stability based on cross-validation. Authors demonstrate the asymptotic selection consistency of their procedure. However, the expected value of this measure is related to the number of individuals. Consequently, since by data splitting cross-validation reduces the sample size, the stability estimate could be biased. For this reason, \citet{Fang12} proposed an insightful bootstrap technique avoiding data splitting.

In this paper, we generally more focus on the pooling step, both in terms of partitions pooling and in terms of sample variability with incomplete data. The rest of the paper is as follows. Based on the literature on consensus clustering and stability assessment in the context of complete data, we argue in Section 2 how to apply Rubin's rules after MI. Then in Section 3, our methodology is assessed by simulation. Finally, an application to a real data set is proposed to determine the number of clusters when data are incomplete.

\section{Method}
\subsection{Notations}
Following standard notations for incomplete data analysis, we denote $\bfX=\left(x_{\nbind\nbvar}\right)_{1\leq \nbind\leq \Nbind,1\leq \nbvar\leq \Nbvar}$ the full data set and $\bfR=\left(r_{\nbind\nbvar}\right)_{1\leq \nbind\leq \Nbind,1\leq \nbvar\leq \Nbvar}$ the missing data pattern, so that $r_{\nbind\nbvar}=0$ if $x_{\nbind\nbvar}$ is missing and 1 if observed. $X$ and $R$ are the associated random variables. For a given observation $\nbind$, the set of observed values is denoted $x_\nbind^{obs}$, while the set of missing values is denoted $x_\nbind^{miss}$, so that $x_\nbind=\left(x_\nbind^{obs},x_\nbind^{miss}\right)$. Note that this partition of variables is specific for each observation. Similarly, we denote $X^{obs}$ and $X^{miss}$ the observed and the missing part of $X$ so that $X=\left(X^{obs},X^{miss}\right)$. The distribution of random variable is denoted $F$. Next, we place ourselves under the standard missing at random (MAR) assumption, meaning $R$ and $X^{miss}$ are independent conditionally to $X^{obs}$ \cite{Rubin76}.

\subsection{Rubin's rules}

From a frequentist point of view, MI aims at estimating the expected mean and the expected variance of a statistic $\QI$ over the realizations of $\left(X^{obs},R\right)$. For instance, $\QI$ could be the least squared estimator of regression coefficients in a linear model, or the estimator of correlation between variables, etc. Under the MAR assumption, consistent estimators can be obtained by ignoring the distribution of $R$. The associated estimates are obtained in three steps:
 \begin{enumerate}
     \item $\Nbtab$ values of $X^{miss}$ are drawn from their predictive distribution $F_{X^{miss}\vert X^{obs}}$, leading to $\Nbtab$ imputed data sets $\left(\bfX^{obs},\bfX^{miss}_\nbtab \right)_{1\leq \nbtab\leq \Nbtab}$.
     \item $\QI$ is evaluated on each one, providing a set of point estimates $\left(\hat{\QI}_\nbtab\right)_{1\leq\nbtab\leq \Nbtab}$ with $\hat{\QI}_\nbtab=Q\left(\bfX^{obs},\bfX^{miss}_\nbtab\right)$, as well as a set of estimated variances $\left(U_\nbtab\right)_{1\leq\nbtab\leq \Nbtab}$.
     \item These values are aggregated according to the Rubin's rules \citep{Rubin76} leading to a unique point estimate $\bar{\QI}$ and a unique variance estimate $T$ as follows:
     \begin{equation}
 \bar{\QI}=\frac{1}{\Nbtab}\sum_{\nbtab=1}^\Nbtab \widehat{\QI}_\nbtab \label{rubin1}
 \end{equation}

\begin{eqnarray}
 T&=&\underbrace{\frac{1}{\Nbtab}\sum_{\nbtab=1}^\Nbtab U_\nbtab}_{\bar{U}}+\underbrace{\frac{1}{\Nbtab-1}\sum_{\nbtab=1}^\Nbtab \left(\widehat{\QI}_\nbtab- \bar{\QI}\right)^2}_B\label{rubin2}
 \end{eqnarray}
 \end{enumerate}

By the second step, we obtain $\Nbtab$ independent realizations of $\QI$ given $X^{obs}$. These realizations are centered around $\QI\left(\bfX^{obs},\bfX^{miss}\right)$ (in expectation), \textit{i.e.} around the value of the statistic which would be observed if data were fully observed. Consequently, their average given by $\bar{\QI}$ in the third step is an unbiased estimate of the expected value of $\QI$ over all observed samples. Thus, $\bar{Q}$ estimates
\begin{equation}
   \mathbb{E}_{X^{obs}}\left[\mathbb{E}_{X}\left[\QI\left(X^{obs},X^{miss}\right)\vert X^{obs}\right]\right]
   \label{form1}
\end{equation}

Similarly, $\bar{U}=\frac{1}{\Nbtab}\sum_{\nbtab=1}^\Nbtab U_\nbtab$ estimates the variance of $\QI$ over observed samples and\\ $B=\frac{1}{\Nbtab-1}\sum_{\nbtab=1}^\Nbtab \left(\widehat{\QI}_\nbtab- \bar{\QI}\right)^2$ the additional variance related to missing values. Following an ANOVA decomposition, the total variance $T$ is expressed as the sum of a within imputation variance ($\bar{U}$) and a between imputation variance ($B$), so that $T$ estimates 
\begin{eqnarray}
     \underbrace{\mathbb{E}_{X^{obs}}\left[Var_{X}\left[\QI\left(X^{obs},X^{miss}\right)\vert X^{obs}\right]\right]}_{within} 
     + \underbrace{Var_{X^{obs}}\left[\mathbb{E}_{X}\left[\QI\left(X^{obs},X^{miss}\right)\vert X^{obs}\right]\right]}_{between}
     \label{form2}
\end{eqnarray}

Note that according to the values of $\left(\bfX^{miss}_\nbtab\right)_{1\leq \nbtab\leq \Nbtab}$, $\bar{\QI}$ randomly varies around its expectation given $X^{obs}$. When $\Nbtab$ is small, this additional variability cannot be ignored
. For this reason, $B$ is generally corrected to account for it. Such a correction is obtained by multiplying $B$ by $(1+1/\Nbtab)$ \citep{Schafer97}.
 
Compared to single imputation, MI accounts for the variability due to missing values thanks to the between variance $B$
. The ratio $B/T$ is helpful for interpretation since it assesses the robustness of the final analysis results to the imputation model \citep{VB18}.

\paragraph{Challenges and motivations}

Rubin's rules (Equations \ref{rubin1} and \ref{rubin2}) have been developed for statistic $\QI$ in $\mathbb{R}$ \citep{Marshall09}. However, a clustering algorithm does not lie within this scope, since it can be expressed as a categorical variable $\Psi$ with values in the set of partitions of $\Nbind$ observations in $\Nbgroup$ clusters at the most. Thus, this paper aims to develop new rules in the context of cluster analysis: a first rule to pool the $\Nbtab$ realizations of $\Psi$ obtained from each imputed data set, as well as a second rule to compute a unique associated uncertainty measure which accounts for sample variability and missing values. Such an innovative methodology would offer a new way for applying any cluster analysis method on incomplete data.

\subsection{Partitions pooling}
In the context of clustering, partitions pooling refers to consensus clustering. Based on the literature in this research field when data are complete, we are going to propose an equivalent of the first rule (Equation \ref{rubin1}).
 
 \subsubsection{Consensus with complete data}
Consensus clustering has been addressed by several authors since several decades (see e.g. \citet{Day86, Vega11} for a survey). The main idea of consensus clustering is to agglomerate the separate partitions $\left(\Psi_\nbpart\right)_{1\leq \nbpart\leq \Nbpart}$ (called \textit{contributory partitions}) into a global partition that must be as similar as possible to the contributory partitions according to an index, like the Rand index defined as the proportion of agreements between partitions, \textit{\textit{i.e.}} $\frac{1}{\Nbind(\Nbind-1)}\sum_{\left(\nbind,\nbind'\right)}\delta_{\nbind\nbind'}$ where $\delta_{\nbind\nbind'}$ is equal to 0 if individuals $\nbind$ and $\nbind'$ belong to the same cluster in one partition and not in the other; and $\delta_{\nbind\nbind'}$ is equal to 1 otherwise. These contributory partitions can be due to various algorithms, or several tuning parameters, several sets of features etc. Recently, \citet{Jain17} brought a strong theoretical framework to consensus clustering by seeing a consensus algorithm as an estimate of the partition minimizing the expected sum of the dissimilarity $\delta$ with all separate partitions (over all partitions). More precisely, the expected partition is defined as follows
\begin{equation}
    argmin_{\Psi\in \mathcal{P}_{\Nbind,\Nbgroup}} \int_{\mathcal{P}_{\Nbind,\Nbgroup}}\delta^{\alpha}(\Psi^\star,\Psi)d\pi(\Psi^\star)\label{eqn:meanpart}
\end{equation}
where $\pi$ is a probability distribution on the partition space $\mathcal{P}_{\Nbind,\Nbgroup}$ of $\Nbind$ observations in $\Nbgroup$ clusters at the most, $\delta$ a dissimilarity function and $\alpha$ a positive real. An estimator of \eqref{eqn:meanpart} can be defined as the partition minimizing the loss function from the observed contributory partitions $\left(\Psi_\nbpart\right)_{1\leq \nbpart\leq \Nbpart}$:

\begin{equation}
 L(\Psi)=\sum_{\nbpart=1}^\Nbpart \delta^{\alpha}(\Psi,\Psi_\nbpart)\label{critcons}
\end{equation}

By this theoretical framework, \citet{Jain17} extends the notion of mean (dedicated to real statistic) to the context of partitions.

However, because of the huge number of partitions, minimizing (\ref{critcons}) is highly challenging. The literature mainly deals with $\alpha$ tuned to 1 or 2, referred as \textit{median partition problem}. \citet{Vega11} distinguished four families of methods for solving it: non-negative matrix factorization (NMF) based methods, Mirkin distance-based methods, Kernel methods and genetics algorithms. The first two have interesting theoretical properties.

NMF methods consists in rewriting the median partition problem as
\begin{equation}
    arg min_{\bfH}\parallel \bfM-\bfH\parallel^2
\end{equation}
where $\parallel . \parallel$ denotes the Frobenius norm, $\bfH$ $\left(\Nbind \times \Nbind\right)$ denotes a connectivity matrix (meaning $\bfH=(h_{\nbind\nbind'})_{1\leq \nbind,\nbind'\leq \Nbind}$ with $h_{\nbind\nbind'}=1$ if the individuals $\nbind$ and $\nbind'$ are in a same cluster and $h_{\nbind\nbind'}=0$ otherwise) and $\bfM=\frac{1}{\Nbpart}\sum_{\nbpart=1}^\Nbpart \bfH_\nbpart$ denotes
the mean of the connectivity matrices $\left(H_\nbpart\right)_{1\leq \nbpart\leq \Nbpart}$ associated to each contributory partition $\Psi_\nbpart$ ($1\leq \nbpart\leq \Nbpart$). The constraint that $\bfH$ must be a connectivity matrix can be expressed as an optimization over the set of the orthogonal matrices \citep{Li07}. The solution of this problem under orthogonality constraint corresponds to the partition minimizing the loss function given in Equation (\ref{critcons}). NMF is a powerful method widely used for solving many optimization problems (beyond the clustering framework) in several fields. One of the main theoretical strengths of the method is the monotone convergence of the optimization algorithm used.

Mirkin distance-based methods focus on minimizing the loss function (\ref{critcons}) for $\delta$ chosen as the number of disagreements between partitions (called \textit{Mirkin distance}). The Mirkin distance does not make the problem less complex, but it has been widely studied and benefits from theoretical results. For example, when the solution is restricted to the set of contributory partitions, the error cannot exceed two times the one obtained by the global optimum \citep{FILKOV04}.



Note that many other methods have been proposed to perform consensus clustering, like the popular Cluster based Similarity Partitioning Algorithm (CSPA), which consists in re-clustering the individuals from the average ($\bfM$) of the connectivity matrices associated to the contributory partitions. However, those methods cannot be expressed as a median partition problem and consequently cannot be justified from an inferential point of view. See \citet{Vega11,Strehl02} for a review.

\subsubsection{Partitions pooling after MI}

Partitions pooling after MI aims at aggregating several partitions varying by the imputed values only. As the first Rubin's rule is motivated by inferential argument, consensus clustering based on the median partition problem is theoretically appealing since it directly extends the notion of expected mean to the clustering framework, providing a straightforward application of the first Rubin's rule to clustering. Among the consensus methods based on the median partition problem, genetics algorithms do not offer theoretical guaranties, while kernel-based methods seem irrelevant in the context of MI. Indeed, since imputed values are generated independently from their predictive distribution, we assume all contributory partitions should have the same weight. Thus, only NMF-based methods and Mirkin distance-based methods provide a suitable way to pool partitions after MI. Formally, our equivalent of the first Rubin's rule for clustering after MI is as follows:
\begin{equation}
\bar{\Psi}=argmin_{\Psi} \sum_{\nbtab=1}^\Nbtab \delta(\Psi,\Psi_\nbtab)\label{newrule1}
\end{equation}
with $\Psi_\nbtab$ is the partition obtained from $\left(\bfX^{obs},\bfX^{miss}_\nbtab\right)$ and $\delta$ the Mirkin distance, or equivalently

\begin{equation}
    arg min_{\bfH}\parallel \bfM-\bfH\parallel^2 \label{newrule1bis}
\end{equation}
with $\bfM=\frac{1}{\Nbtab}\sum_{\nbtab=1}^\Nbtab \bfH_\nbtab$ and $\bfH_\nbtab$ the connectivity matrix associated to $\Psi_\nbtab$.

Following \citet{Jain17}, the obtained partition estimates the theoretical partition minimizing the expected sum of the dissimilarity $\delta$ with all separate partitions given $X^{obs}$.

Based on other techniques, clustering after MI has been previously investigated. Some authors have proposed stacking centroids or stacking imputed data sets for dealing with several imputed data sets \citep{Plaehn19}. We can note that stacking ignores the differences between imputed data sets. Among other methods, \citet{Basagana13} proposed a general framework for consensus by previously investigating the choice of the number of partitions and the choice of the subset of variables retained for clustering. For a given number of clusters and a set of variables, consensus is performed by majority vote over the contributory partitions. More recently, \citet{Bruckers17} proposed a consensus for functional data. For achieving this goal, they first identify the indicator matrices associated to each contributory partition and then, they look at the fuzzy matrix minimizing the Euclidean distance over the indicator matrices. The consensus partition is then obtained by majority vote \citep{Dimitriadou02}. Like CSPA, this approach in two steps cannot be considered as based on the median partition problem \citep{Vega11}. Lately, \citet{Faucheux} proposed consensus based on the MultiCons algorithm \citep{Al-Najdi16}. The algorithm presents many advantages, in particular it allows a visualization of the hidden cluster structure in the data set, but it does not aim at minimizing the median partition problem \citep[p. 16]{Al-Najdi16}.

While these other methods yield a unique partition from several imputed data sets, this partition cannot be expressed as an estimator of a theoretical partition, contrary to those based on the median partition problem.

\subsection{Instability pooling}
Based on the literature in cluster stability when data are complete, we now propose an equivalent of the second rule (Equation \ref{rubin2}).

\subsubsection{Instability with complete data}
Assessing the instability in clustering is important for data analysis. For achieving this goal, resampling methods are appealing, especially when the clustering algorithm is distance-based, like k-means, k-medoids or hierarchical clustering. \citet{Wang10} and \citet{Fang12} proposed two ways for computing instability measure from any clustering algorithms. The first one is based on cross-validation, while the second is based on bootstrap. Since the cross-validation method tends to underestimate the instability \citep{Wang10,Fang12}, the bootstrap method appears like more relevant. The main idea consists in defining a theoretical distance (instability) $\delta$ between clusterings based on the sample distribution $F_X$. Then, this distribution is mimicked by bootstrap and the distance is evaluated from each bootstrap replicate. Finally, distances are aggregated by averaging. More precisely, the theoretical distance $\delta_{F_X}$ between two clusterings $\Psi$ and $\Psi'$ is defined by
\begin{equation}
\delta_{F_X}(\Psi_\nbpart,\Psi_{\nbpart'})=\mathbb{P}_{F_X}\lbrace I\left(V_{\Psi_{\nbpart}}\left(X\right)=V_{\Psi_{\nbpart}}\left(X'\right)\right)+I\left(V_{\Psi_{\nbpart'}}\left(X\right)=V_{\Psi_{\nbpart'}}\left(X'\right)\right)=1\rbrace
\end{equation}

where $X$ and $X'$ are independently drawn from the distribution $F_X$ and $V_{\Psi_{\nbpart}}(X)$ is the Voronoi cell for $X$ according to the partition given by $\Psi_{\nbpart}$.
This distance measures the probability of disagreement between both clusterings.

Based on this definition, the instability of $\Psi$ is defined as 
\begin{equation}
\mathbb{E}_{X^\Nbind\sim F_X^\Nbind,\tilde{X}^\Nbind\sim F_X^\Nbind}\left[\delta_{F_X}\left(\Psi\left(X^\Nbind\right),\Psi\left(\tilde{X}^\Nbind\right)\right)\right]\label{instability}
\end{equation}
\textit{i.e.} as the expectation over all random samples of size $\Nbind$ of the distances between partitions given by clustering trained on them.
 
\citet{Fang12} proposes an estimate of (\ref{instability}) by bootstrap: $\Nbboot$ bootstrap pairs $\left(\bfX_{\nbboot},\tilde{\bfX}_{\nbboot}\right)_{1\leq \nbboot\leq\Nbboot}$ are drawn from the empirical distribution $\widehat{F}^\Nbind$. From each one, $\Psi$ is evaluated. Both estimates are used to classify the individuals of $\bfX$ according to the Voronoi cells defined by each partition (e.g. by considering the closest centroid). Finally, the instability of the clustering is estimated by
\begin{equation}
U^{boot}= \frac{1}{\Nbboot}\sum_{\nbboot=1}^\Nbboot \delta_{\hat{F}^\Nbind}\left(\Psi\left(\bfX_\nbboot\right),\Psi\left(\tilde{\bfX}_\nbboot\right)\right)\label{instabilite}
\end{equation}
with 
\begin{eqnarray}
\delta_{\hat{F}^\Nbind}\left(\Psi\left(\bfX_\nbboot\right),\Psi\left(\tilde{\bfX}_\nbboot\right)\right)&=&\frac{1}{\Nbind^2}\sum_{\nbind=1}^\Nbind\sum_{\nbind'=1}^\Nbind \vert I\left(V_{\Psi\left({\bfX}_\nbboot\right)}\left(x_\nbind\right)=V_{\Psi\left({\bfX}_\nbboot\right)}\left(x_{\nbind'}\right)\right)\nonumber\\&&-I\left(V_{\Psi\left(\tilde{\bfX}_\nbboot\right)}\left(x_\nbind\right)=V_{\Psi\left(\tilde{\bfX}_\nbboot\right)}\left(x_{\nbind'}\right)\right)\vert\label{calculMirkin}
\end{eqnarray}

Note that $\delta_{\hat{F}^\Nbind}$ corresponds to the proportion of disagreements between both partitions of $\bfX$ (up to the scalar factor $\frac{\Nbind}{\Nbind-1}$) and can be viewed as a normalized Mirkin distance. \citet{Fang12} indicate moderate values of $\Nbboot$ (20 or 50) are sufficient for a precise instability assessment. Furthermore, this instability can be assessed for distance-based or non-distance-based clustering algorithms.
\subsubsection{Instability with incomplete data \label{sectioninst}}

Following previous developments for complete data, the instability with missing data can be defined as the expectation given in (\ref{instability}) over observed data. Following the second Rubin's rule, such an instability can be decomposed as the sum of a within instability (Equation \ref{Rubinincompleteintra}) and a between instability (Equation \ref{Rubinincompleteextra}):
{\begin{equation}
\mathbb{E}_{X^{obs},\tilde{X}^{obs}}\left[\mathbb{E}_{X,\tilde{X}
}\left[\delta_{F_{X\vert X^{obs}}}\left(\Psi\left(X^{obs}, X^{miss}\right),\Psi\left(\tilde{X}^{obs}, \tilde{X}^{miss}\right)\right)\vert X^{obs}\right]\right]\label{Rubinincompleteintra}
\end{equation}}

\begin{equation}
\mathbb{E}_{
X^{obs},
\tilde{X}^{obs}
}\left[\delta_{F_{X}}\left(\Psi\left(X^{obs}, X^{miss}\right),\Psi\left(\tilde{X}^{obs}, \tilde{X}^{miss}\right)\right)\vert X^{obs}\right]
\label{Rubinincompleteextra}
\end{equation}

\normalsize
Following Equation (\ref{rubin2}), the within instability (Equation \ref{Rubinincompleteintra}) can be estimated
from imputed $\Nbtab$ data sets $\left(\bfX^{obs},\bfX^{miss}_\nbtab\right)_{1\leq \nbtab \leq\Nbtab}$ by: 
\begin{equation}
 \bar{U}=\frac{1}{\Nbtab}\sum_{\nbtab=1}^\Nbtab U_\nbtab^{boot}\label{intra}
 \end{equation}
where $U_\nbtab^{boot}$ is the instability estimated from $\left(\bfX^{obs},\bfX^{miss}_\nbtab\right)$ according to Equation (\ref{instabilite}) and the between instability (Equation \ref{Rubinincompleteextra}) can be estimated by:
\begin{equation}
B=\frac{1}{\Nbtab^2} \sum_{\nbtab=1}^\Nbtab\sum_{\nbtab'=1}^\Nbtab \delta_{\hat{F}_{X\vert X^{obs}}}\left(\Psi\left(\bfX^{obs},\bfX^{miss}_\nbtab\right),\Psi\left(\bfX^{obs},\bfX^{miss}_{\nbtab'}\right)\right)
\label{B3} 
\end{equation} 
where $\delta_{\hat{F}_{X\vert X^{obs}}}\left(\Psi\left(\bfX^{obs},\bfX^{miss}_\nbtab\right),\Psi\left(\bfX^{obs},\bfX^{miss}_{\nbtab'}\right)\right)$ corresponds to the proportion of disagreements between partitions obtained from imputed data sets $\nbtab$ and $\nbtab'$ as in Equation \eqref{calculMirkin}. We note this expression does not depend on the mean partition, while the rule in Equation (\ref{rubin2}) depends on the mean. For this reason, no correction for small values of $\Nbtab$ is required here.

The total instability is given by the sum of Equations (\ref{intra}) and (\ref{B3}):
\begin{equation}
    T=\frac{1}{\Nbtab}\sum_{\nbtab=1}^\Nbtab U_\nbtab^{boot}+\frac{1}{\Nbtab^2} \sum_{\nbtab=1}^\Nbtab\sum_{\nbtab'=1}^\Nbtab \delta_{\hat{F}_{X\vert X^{obs}}}\left(\Psi\left(\bfX^{obs},\bfX^{miss}_\nbtab\right),\Psi\left(\bfX^{obs},\bfX^{miss}_{\nbtab'}\right)\right)\label{total}
\end{equation}

Note that without missing values, Equation (\ref{total}) is equivalent to the instability proposed in \citet{Fang12}. More generally, $T$ is positive and bounded by 2. The value of 0 is reached if $\Nbgroup=1$ or if the clustering is constant whatever the incomplete sample. The less stable the clustering is, the larger $T$ will be. A large value of the between instability compared to the total one indicates a strong dependence of the clustering to the imputation model.

\subsection{Summary}
Based on above developments, the full procedure to perform cluster analysis after multiple imputation can be summarized as follows:
\begin{description}
     \item[\textbf{Imputation}] From an incomplete data set, generate $\Nbtab$ imputed data sets according to a predefined multiple imputation method
     \item[\textbf{Analysis}] For $\nbtab$ in  $\lbrace 1\ldots\Nbtab\rbrace$:
     \begin{enumerate}
         \item build a partition $\Psi_\nbtab$ from the $\nbtab^{th}$ imputed data set
         \item compute $U_\nbtab^{boot}$ the associated instability:
         \begin{enumerate}
             \item by resampling individuals with replacement, generate $\Nbboot$ bootstrap pairs $\left(\bfX_{\nbboot},\tilde{\bfX}_{\nbboot}\right)_{1\leq \nbboot\leq\Nbboot}$ from the $\nbtab^{th}$ imputed data set
                \item for each bootstrap pair $\nbboot$ in $\lbrace 1\ldots\Nbboot\rbrace$
                \begin{itemize} 
                \item perform cluster analysis from $\left(\bfX_{\nbboot},\tilde{\bfX}_{\nbboot}\right)_{1\leq \nbboot\leq\Nbboot}$ to obtain a pair of partitions $\left(\Psi_\nbboot,\tilde{\Psi}_\nbboot\right)$
                \item classify the $\Nbind$ individuals of the $\nbtab^{th}$ imputed data set according to $\Psi_\nbboot$ and $\tilde{\Psi}_\nbboot$ to obtain a pair of partitions $\left(\Psi'_\nbboot,\tilde{\Psi}'_\nbboot\right)$
                \item compute the proportion of disagreements between $\Psi'_\nbboot$ and $\tilde{\Psi}'_\nbboot$ as  $U_\nbtab^\nbboot=\frac{1}{\Nbind^2}\sum_{\left(\nbind,\nbind'\right)}\delta_{\nbind,\nbind'}$  where $\delta_{\nbind\nbind'}$ is equal to 0 if individuals $\nbind$ and $\nbind'$ belong to the same cluster in one partition and not in the other; and $\delta_{\nbind\nbind'}$ is equal to 1 otherwise
                \end{itemize}
                \item compute the instability associated to $\Psi_\nbtab$ by averaging $U_\nbtab^{boot}=\frac{1}{\Nbboot}\sum_{\nbboot=1}^{\Nbboot}U_\nbtab^\nbboot$
         \end{enumerate}
     \end{enumerate}
     \item[\textbf{Pooling}] The set of partitions $\left(\Psi_\nbtab\right)_{1\leq \nbtab\leq\Nbtab}$ and the set of associated instability estimates $\left(U_\nbtab^{boot}\right)_{1\leq \nbtab\leq\Nbtab}$ are aggregated as follows:
     \begin{description}
         \item[\textbf{First rule (partitions pooling)}]  using NMF or Mirkin based methods, compute the consensus partition as \begin{equation*}\bar{\Psi}=argmin_{\Psi} \sum_{\nbtab=1}^\Nbtab \delta(\Psi,\Psi_\nbtab)\end{equation*} where $\delta(\Psi,\Psi_\nbtab)$ denotes the number of disagreements between partitions $\Psi$ and $\Psi_\nbtab$ (Mirkin distance)
         \item[\textbf{Second rule (instability pooling)}] compute the total instability as:
\begin{equation*} T=\frac{1}{\Nbtab}\sum_{\nbtab=1}^\Nbtab U_\nbtab^{boot}+\frac{1}{\Nbtab^2} \sum_{\nbtab=1}^\Nbtab\sum_{\nbtab'=1}^\Nbtab \delta\left({\Psi_{\nbtab}},{\Psi_{\nbtab'}}\right)/\Nbind^2
\end{equation*}
     \end{description}
\end{description}
Note that an implementation is provided through an R package entitled \textit{clusterMI} available at the web page\footnote{\url{https://vincentaudigier.weebly.com/}} of the first author.

\section{Simulations}
After proposing rules for pooling clusterings after MI, we highlight how the pooled results vary according to the data structure. Furthermore, we investigate their robustness to the number of imputed data sets $\Nbtab$. The R code used for simulations is available on demand.

\subsection{Simulation design}
\subsubsection{Data generation}
Full data are simulated according to a $\Nbvar$ multivariate Gaussian mixture model with two mixture components:
\begin{equation*}
    X\sim \pi_1\mathcal{N}_\Nbvar\left(\mu_1,\Sigma\left(\rho\right)\right)+\pi_2\mathcal{N}_\Nbvar\left(\mu_2,\Sigma\left(\rho\right)\right)
\end{equation*}

where $\Nbvar=10$, $\pi_1=\pi_2=1/2$, $\mu_1=(0,0,0,0,0,0,0,0,0,0)$, $\mu_2=(0,0,0,0,0,2,2,2,2,2)$ and

$\Sigma\left(\rho\right)={\scriptsize{\left(\begin{array}{cc}
    I_5 &  \text{\huge0}\\
    \text{\huge0} & \begin{array}{ccccc}1&\rho&\rho&\rho&\rho\\
    \rho&1&\rho&\rho&\rho\\
    \rho&\rho&1&\rho&\rho\\
    \rho&\rho&\rho&1&\rho\\
    \rho&\rho&\rho&\rho&1
    \end{array}
\end{array}\right)}}$. Several configurations are investigated by varying the number of individuals $\Nbind\in\{25,50,100,200\}$, the correlation between variables $\rho\in\{0.3, 0.6\}$. For each configuration, $\Nbsim=200$ data sets are generated. For each data set, several missing data patterns are considered varying by the percentage of missing values $\tau \in \{0.1, 0.3, 0.5\}$ and their distribution: $Prob(r_{\nbind\nbvar}=0)=\tau$ for all $1\leq\nbind\leq\Nbind$ and $1\leq\nbvar\leq\Nbvar$ (MCAR mechanism) or $Prob(r_{\nbind\nbvar}=0)=\Phi(a_\tau+x_{\nbind 1})$ for all $1\leq\nbind\leq\Nbind$ and $2\leq\nbvar\leq\Nbvar$ with $\Phi$ the cumulative distribution function of the standard normal distribution (MAR mechanism) and $a_\tau$ a constant to control the percentage of missing values in expectation. Thus, 9600 incomplete data sets are investigated. Note the computational cost to investigate each one does not allow a larger number of replications.

\subsubsection{Methods}
Each incomplete data set is imputed according to two MI methods accounting for the structure of individuals \citep{Schafer03}
\begin{itemize}
    \item JM-DP: MI using a non-parametric extension of the mixture model namely the Dirichlet process mixture of products of multivariate normal distributions \citep{Kim14}. The number of components is bounded by 5. The number of iterations for the burn-in period is tuned to 500 and the number of skipped iterations to keep one imputed data set after the burn-in period is tuned to 100.
    \item FCS-RF: MI by random forest \citep{Doove14}. The number of iterations for the multivariate imputation by chained equations algorithm is tuned to 10.
\end{itemize}

For each method, the number of imputed data sets $\Nbtab$ varies in $\{1,5,10,20,50\}$. Note that the case $\Nbtab=1$ corresponds to single imputation. Then, k-means clustering is performed on each imputed data set (using 2 clusters, standardization of variables and 100 initializations) and partitions are pooled according to the proposed rules. 
More precisely, the mean partition is estimated using the NMF clustering based method (Equation \eqref{newrule1}) as proposed in \citet{Li07} and also using a Mirkin distance-based method (Equation \eqref{newrule1bis}) called Simulated-Annealing-One-element-Move (SAOM) \citep{FILKOV04}. Furthermore, the total instability is computed according to Equation \eqref{total}.

As benchmark, k-means clustering is also performed on the full data and on the complete cases. In addition, k-means through a bagging procedure based on bootstrap \citep{Dudoit03} is investigated on full data, while k-means through the k-pod algorithm \citep{Chi16} is investigated on incomplete data. This later algorithm overcomes missing values in k-means by optimizing the k-means criterion over observed values only. For achieving this goal, a majorization-minimization algorithm is used, consisting in alternating clustering of individuals (by k-means) and imputation of incomplete observations by the coordinates of their associated centroid.

\subsubsection{Criteria}
The accuracy of the consensus partition is assessed according to the mean (over the $\Nbsim$ generated data sets) of the adjusted rand index (ARI) \citep{Hubert85} between the consensus partition and the reference one known by simulation.
The mean (over the $\Nbsim$ generated data sets) of the intra instability $\bar{U}$, the mean of the between variability $B$, and the mean of the total instability $T$ are reported for each configuration.

\subsection{Results}

\subsubsection{Partitions pooling}

\begin{figure}
         \begin{subfigure}[b]{.5\linewidth}
    \centering
        \includegraphics[trim={.4cm .7cm 1.6cm 1cm}, clip,height=10cm,width=7cm]{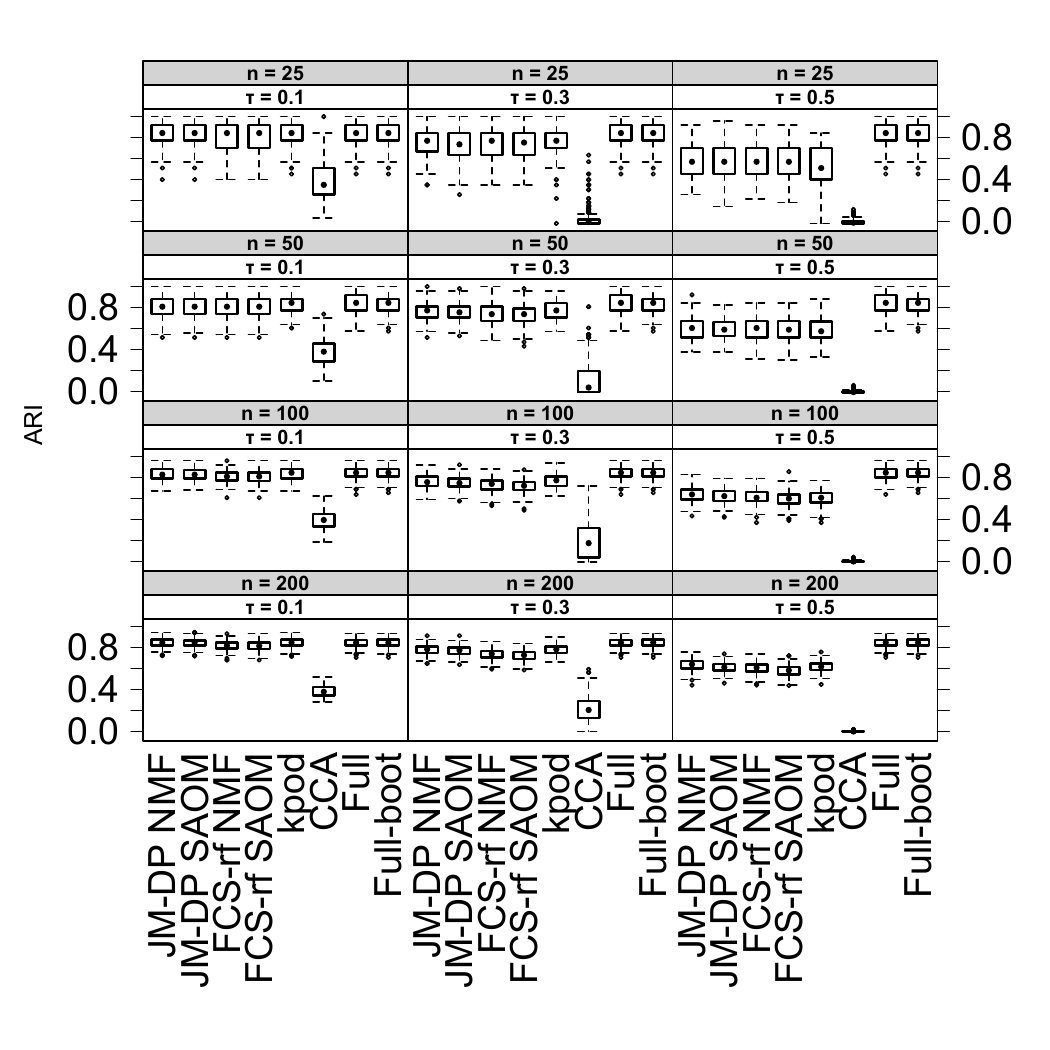}
    \caption{ARI distributions when $\rho=0.3$}
     \end{subfigure}\begin{subfigure}[b]{.48\linewidth}
    \centering
        \includegraphics[trim={2.2cm .7cm .65cm 1cm},clip,height=10cm,width=7cm]{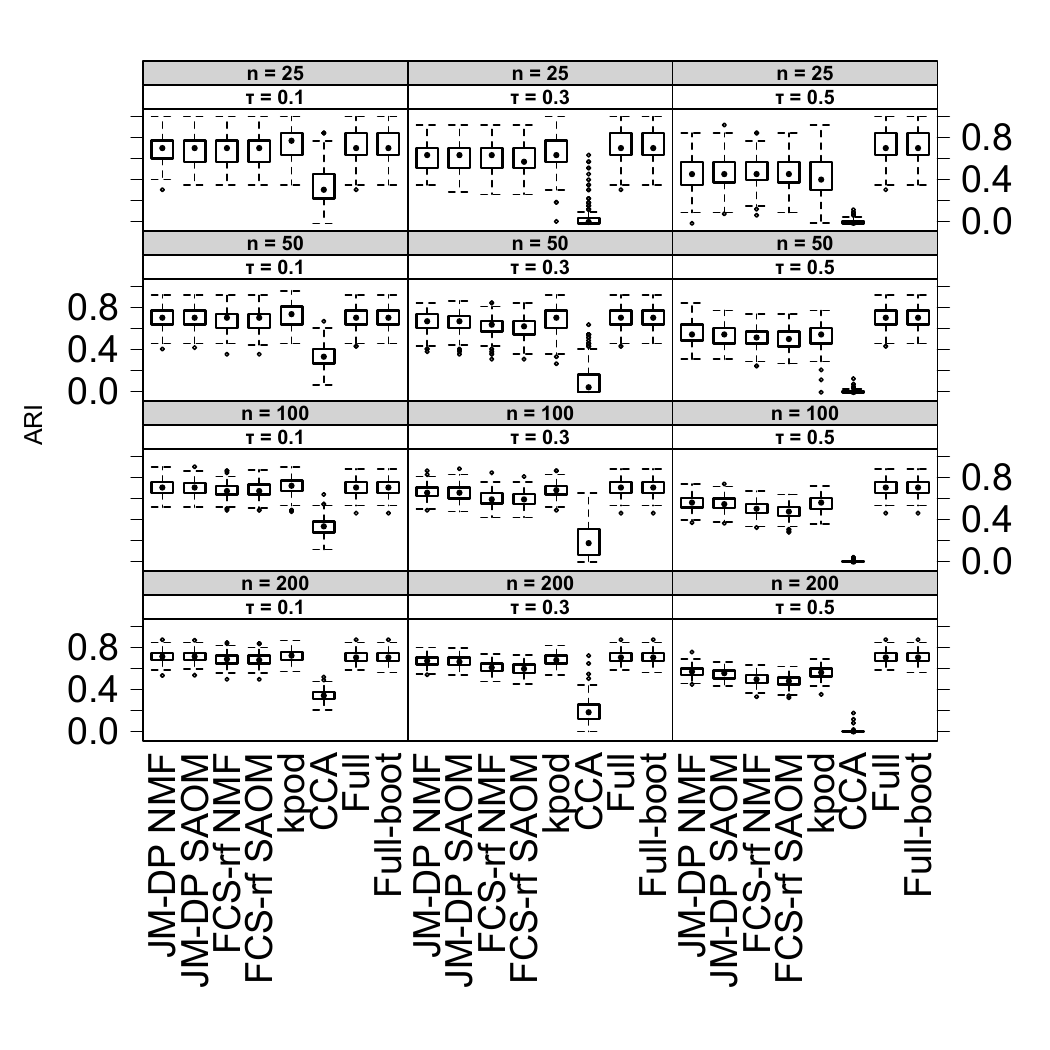}
    \caption{ARI distributions when $\rho=0.6$}
     \end{subfigure}
     \caption{Accuracy of the clustering procedure under a MCAR mechanism: distribution of the adjusted rand index over the $\Nbsim=200$ generated data sets varying by the number of individuals ($\Nbind$), the correlation between variables ($\rho$) and the proportion of missing values ($\tau$) for various imputation methods (JM-DP or FCS-RF), various consensus methods (NMF or SAOM). For each case, clustering is performed using k-means clustering. As benchmark, ARI obtained by applying k-means on complete-cases (CCA), using k-pod algorithm (kpod), on full data (Full) or using a bagging procedure (Full-boot) are also reported.\label{fig:acc_mcar}}
\end{figure}

\begin{figure}
         \begin{subfigure}[b]{.5\linewidth}
    \centering
        \includegraphics[trim={.4cm .7cm 1.6cm 1cm}, clip,height=10cm,width=7cm]{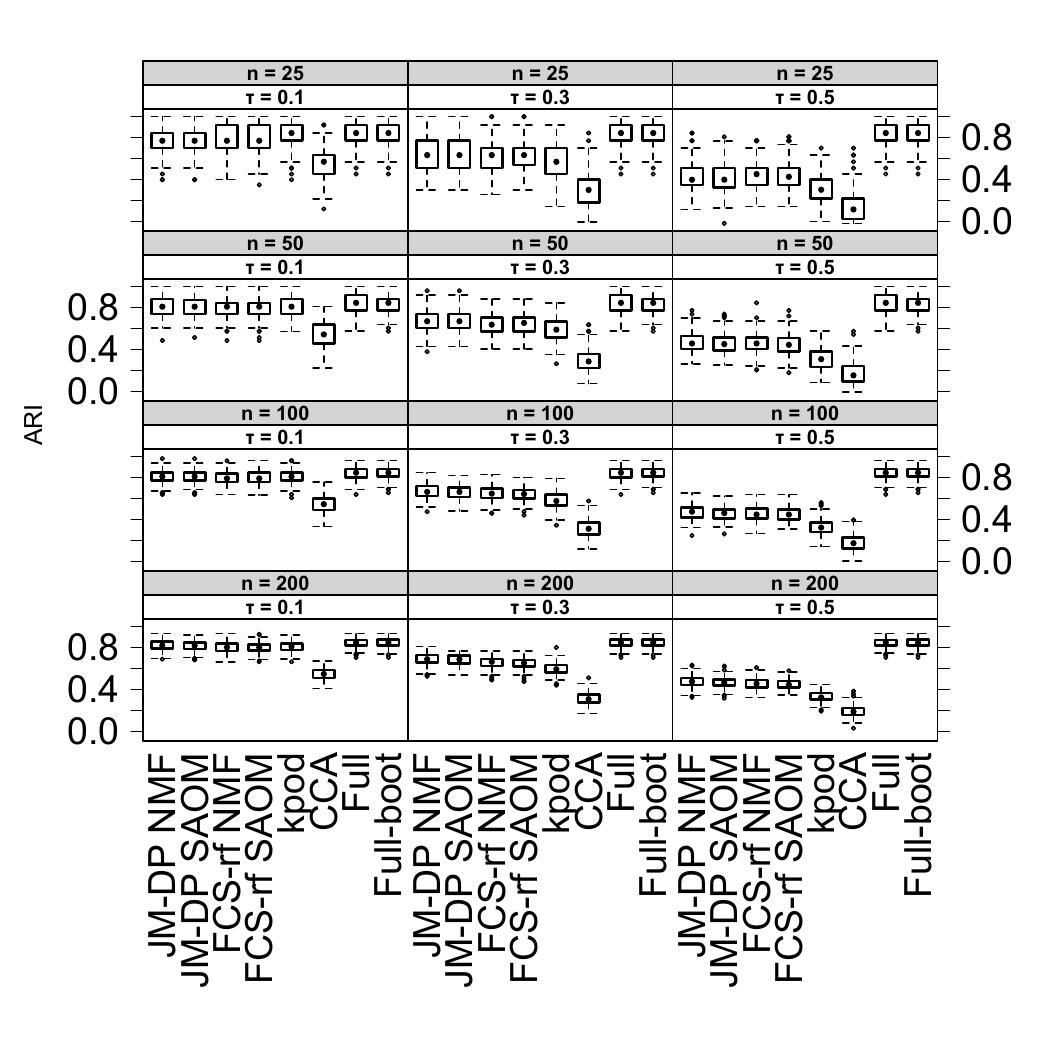}
    \caption{ARI distributions when $\rho=0.3$}
     \end{subfigure}\begin{subfigure}[b]{.48\linewidth}
    \centering
        \includegraphics[trim={2.2cm .7cm .65cm 1cm},clip,height=10cm,width=7cm]{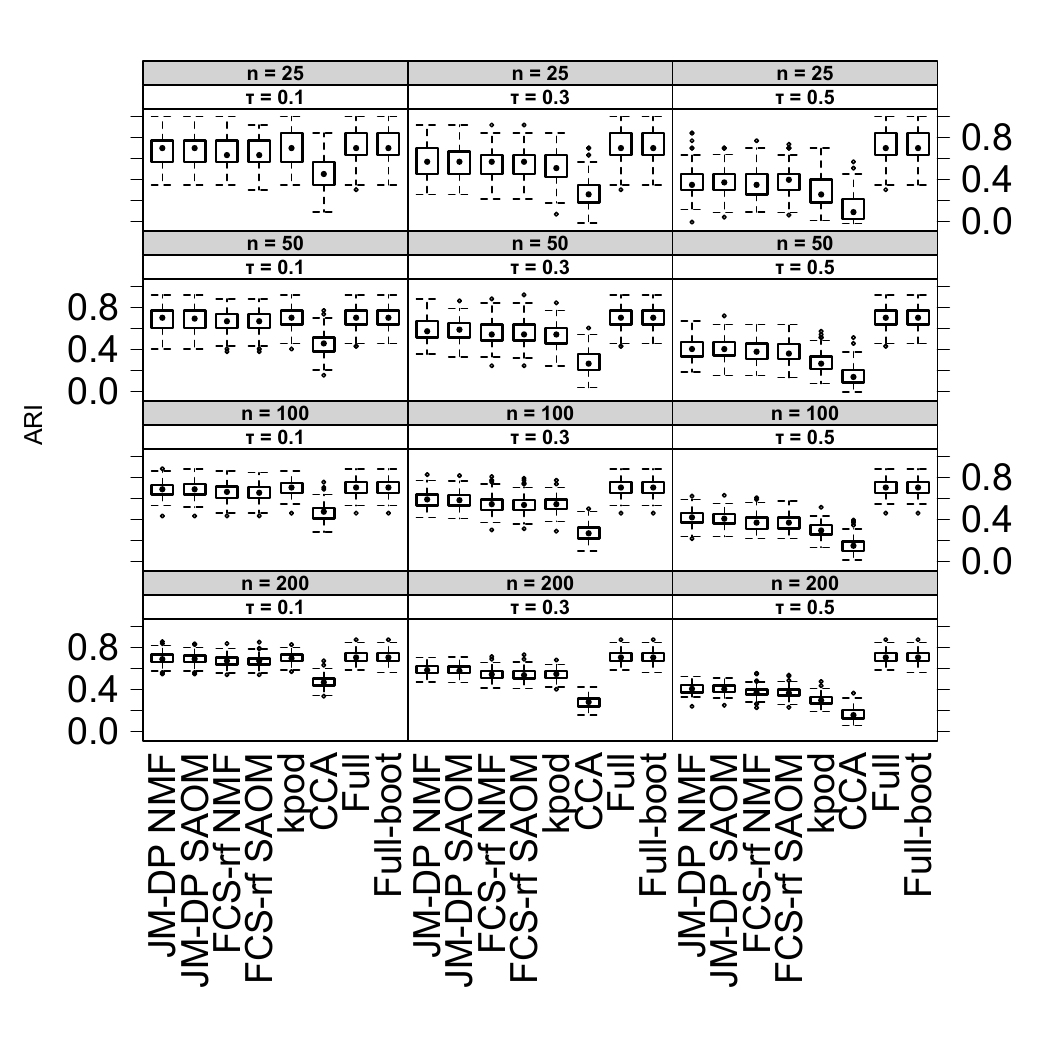}
    \caption{ARI distributions when $\rho=0.6$}
     \end{subfigure}
     \caption{Accuracy of the clustering procedure under a MAR mechanism: distribution of the adjusted rand index over the $\Nbsim=200$ generated data sets varying by the number of individuals ($\Nbind$), the correlation between variables ($\rho$) and the proportion of missing values ($\tau$) for various imputation methods (JM-DP or FCS-RF), various consensus methods (NMF or SAOM). For each case, clustering is performed using k-means clustering. As benchmark, ARI obtained by applying k-means on complete-cases (CCA), using k-pod algorithm (kpod), on full data (Full) or using a bagging procedure (Full-boot) are also reported.\label{fig:acc_mar}}
\end{figure}

Figures \ref{fig:acc_mcar} and \ref{fig:acc_mar} summarize results over the $\Nbsim=200$ simulations for both mechanisms (averages and interquartile ranges are available in Tables \ref{tablearimcar} and \ref{tablearimar} in Appendix). Performances for MI methods consider $\Nbtab=50$ imputed data sets, the influence of $\Nbtab$ is discussed in Section \ref{sec_nbtab}. Note that because complete-case analysis does not cluster all individuals (compared to MI methods or clustering on full data), the following process is applied for a fair comparison: first, complete cases are clustered using kmeans clustering. Then, based on their observed profile, each incomplete case is classified according to the closest centroid. If the data set did not contain complete cases, then a cluster would be assigned at random for each incomplete case. Thus, the resulting partition concerns all observations.  


For both mechanisms, the ARI over the $\Nbsim$ data sets when the contributory partitions are pooled using NMF or when they are pooled using SAOM remain generally close. However, both MI methods show higher ARI values with NMF pooling when the number of individuals and the proportion of missing values increase.



As expected, the ARI obtained by MI is close to the one obtained without missing values (Full or Full-boot) when the proportion of missing values $\tau$ equals 0.1, and the difference is larger when this proportion increases. Furthermore, clustering after MI outperforms complete-cases analysis even if the proportion of missing values is small. Compared to the direct application of kmeans using the k-pod algorithm, similar ARI are observed for a MCAR mechanism, but MI outperforms under the MAR mechanism for a moderate ($\tau=0.3$) or large ($\tau=0.5$) proportion of missing values. 

A large number of individuals slightly increases the ARI and decreases the interquartile range in MI, while it only decreases the interquartile range when data are full.

Finally, the ARI is usually higher when data are imputed using JM-DP than FCS-RF. It could be expected since JM-DP is based on an imputation model close to the model used for data generation, while FCS-RF is based on a non-parametric model.

\subsubsection{Instability pooling}
Table \ref{tableinstmcar} gathers the average of within instability, between instability and total instability over the $\Nbsim=200$ data sets for configurations under a MCAR mechanism with $\Nbind=50$ or $\Nbind=200$ individuals (see Table \ref{tableinstmar} in Appendix for a MAR mechanism and Tables \ref{tableinstnpetitmcar} for $\Nbind\in\{25,100\}$).

\begin{table}
\centering
\caption{Instability of the clustering procedure under a MCAR mechanism: average within-instability ($\bar{U}$) average between-instability ($B$) and average total instability ($T$) over the $\Nbsim=200$ generated data sets for various number of individuals ($\Nbind$), correlation between variables ($\rho$) and proportion of missing values ($\tau$). Two imputation methods are investigated (JM-DP or FCS-RF) using $\Nbtab=1$ or $\Nbtab=50$ imputed data sets. For each case, clustering is performed by k-means. As benchmark, ARI obtained by applying k-means clustering on full data (Full) or complete-case analysis (CCA) are also reported (not possible for a large proportion of missing values). \label{tableinstmcar}} 
\begin{tabular}{llllrrrrrrrr}
  \hline
   \multicolumn{1}{c}{} & \multicolumn{1}{c}{} & \multicolumn{1}{c}{} & \multicolumn{1}{c}{} & \multicolumn{3}{c}{JM-DP} & \multicolumn{3}{c}{FCS-RF} & \multicolumn{1}{c}{Full} & \multicolumn{1}{c}{CCA} \\$\Nbind$ & $\rho$ & $\tau$ & $M$ &  $\bar{U}$ &  $B$ &  $T$ &  $\bar{U}$ &  $B$ &  $T$ & $T$ & $T$ \\ 
  \hline
50 & 0.3 & 0.1 & 1 & 0.08 & 0.00 & 0.08 & 0.07 & 0.00 & 0.07 & 0.06 & 0.13 \\ 
  50 & 0.3 & 0.1 & 50 & 0.08 & 0.06 & 0.14 & 0.07 & 0.06 & 0.13 & 0.06 & 0.13 \\ 
  50 & 0.3 & 0.3 & 1 & 0.13 & 0.00 & 0.13 & 0.11 & 0.00 & 0.11 & 0.06 &  \\ 
  50 & 0.3 & 0.3 & 50 & 0.13 & 0.19 & 0.32 & 0.10 & 0.16 & 0.27 & 0.06 &  \\ 
  50 & 0.3 & 0.5 & 1 & 0.17 & 0.00 & 0.17 & 0.14 & 0.00 & 0.14 & 0.06 &  \\ 
  50 & 0.3 & 0.5 & 50 & 0.17 & 0.37 & 0.54 & 0.15 & 0.31 & 0.45 & 0.06 &  \\ 
  50 & 0.6 & 0.1 & 1 & 0.08 & 0.00 & 0.08 & 0.07 & 0.00 & 0.07 & 0.06 & 0.13 \\ 
  50 & 0.6 & 0.1 & 50 & 0.08 & 0.06 & 0.14 & 0.07 & 0.06 & 0.13 & 0.06 & 0.13 \\ 
  50 & 0.6 & 0.3 & 1 & 0.13 & 0.00 & 0.13 & 0.10 & 0.00 & 0.10 & 0.06 &  \\ 
  50 & 0.6 & 0.3 & 50 & 0.12 & 0.20 & 0.32 & 0.10 & 0.16 & 0.26 & 0.06 &  \\ 
  50 & 0.6 & 0.5 & 1 & 0.17 & 0.00 & 0.17 & 0.14 & 0.00 & 0.14 & 0.06 &  \\ 
  50 & 0.6 & 0.5 & 50 & 0.17 & 0.37 & 0.54 & 0.14 & 0.30 & 0.44 & 0.06 &  \\ 
  200 & 0.3 & 0.1 & 1 & 0.01 & 0.00 & 0.01 & 0.01 & 0.00 & 0.01 & 0.01 & 0.04 \\ 
  200 & 0.3 & 0.1 & 50 & 0.01 & 0.03 & 0.04 & 0.01 & 0.04 & 0.06 & 0.01 & 0.04 \\ 
  200 & 0.3 & 0.3 & 1 & 0.01 & 0.00 & 0.01 & 0.02 & 0.00 & 0.02 & 0.01 & 0.16 \\ 
  200 & 0.3 & 0.3 & 50 & 0.01 & 0.10 & 0.11 & 0.02 & 0.14 & 0.16 & 0.01 & 0.16 \\ 
  200 & 0.3 & 0.5 & 1 & 0.04 & 0.00 & 0.04 & 0.04 & 0.00 & 0.04 & 0.01 &  \\ 
  200 & 0.3 & 0.5 & 50 & 0.04 & 0.25 & 0.29 & 0.04 & 0.26 & 0.30 & 0.01 &  \\ 
  200 & 0.6 & 0.1 & 1 & 0.02 & 0.00 & 0.02 & 0.02 & 0.00 & 0.02 & 0.02 & 0.04 \\ 
  200 & 0.6 & 0.1 & 50 & 0.02 & 0.03 & 0.05 & 0.02 & 0.05 & 0.06 & 0.02 & 0.04 \\ 
  200 & 0.6 & 0.3 & 1 & 0.02 & 0.00 & 0.02 & 0.02 & 0.00 & 0.02 & 0.02 & 0.16 \\ 
  200 & 0.6 & 0.3 & 50 & 0.02 & 0.10 & 0.11 & 0.02 & 0.13 & 0.15 & 0.02 & 0.16 \\ 
  200 & 0.6 & 0.5 & 1 & 0.04 & 0.00 & 0.04 & 0.03 & 0.00 & 0.03 & 0.02 &  \\ 
  200 & 0.6 & 0.5 & 50 & 0.04 & 0.25 & 0.29 & 0.03 & 0.24 & 0.28 & 0.02 &  \\ 
   \hline
  \end{tabular}
\end{table}

As expected, the total instability is always higher by using MI ($\Nbtab=50$) instead of SI ($\Nbtab=1$) for all configurations. Furthermore, clustering instability is generally smaller when using MI instead of complete-case analysis and higher compared to clustering instability when data are full. We note the average between instability ($B$) tends to increase when the proportion of missing values increases. This behavior was expected. More surprisingly, the within instability ($\bar{U}$) is also increasing with the proportion of missing values, even if this increase remains relatively smaller. This is highlighted for small values of $\Nbind$. Such a behavior can be explained by overfitting of the imputation models. Indeed, FCS-RF as non-parametric imputation method requires a large number of observations, while JM-DP as complex model requires a large number of observations to fit accurately the data structure. For this reason, when the number of individuals is small, the imputed values are highly variable, yielding to an increase of the within instability. This behavior is more severe when the proportion of missing values is large.
Note that instability using the k-pod algorithm is not considered since the method only returns a partition from incomplete data, but no instability measure.

\subsubsection{Influence of $\Nbtab$\label{sec_nbtab}}
\begin{figure}
\begin{center}
\begin{subfigure}[t]{.5\textwidth}
\includegraphics[trim={.4cm .7cm 1.6cm 1cm},scale=.45,clip,height=10cm,width=7cm]{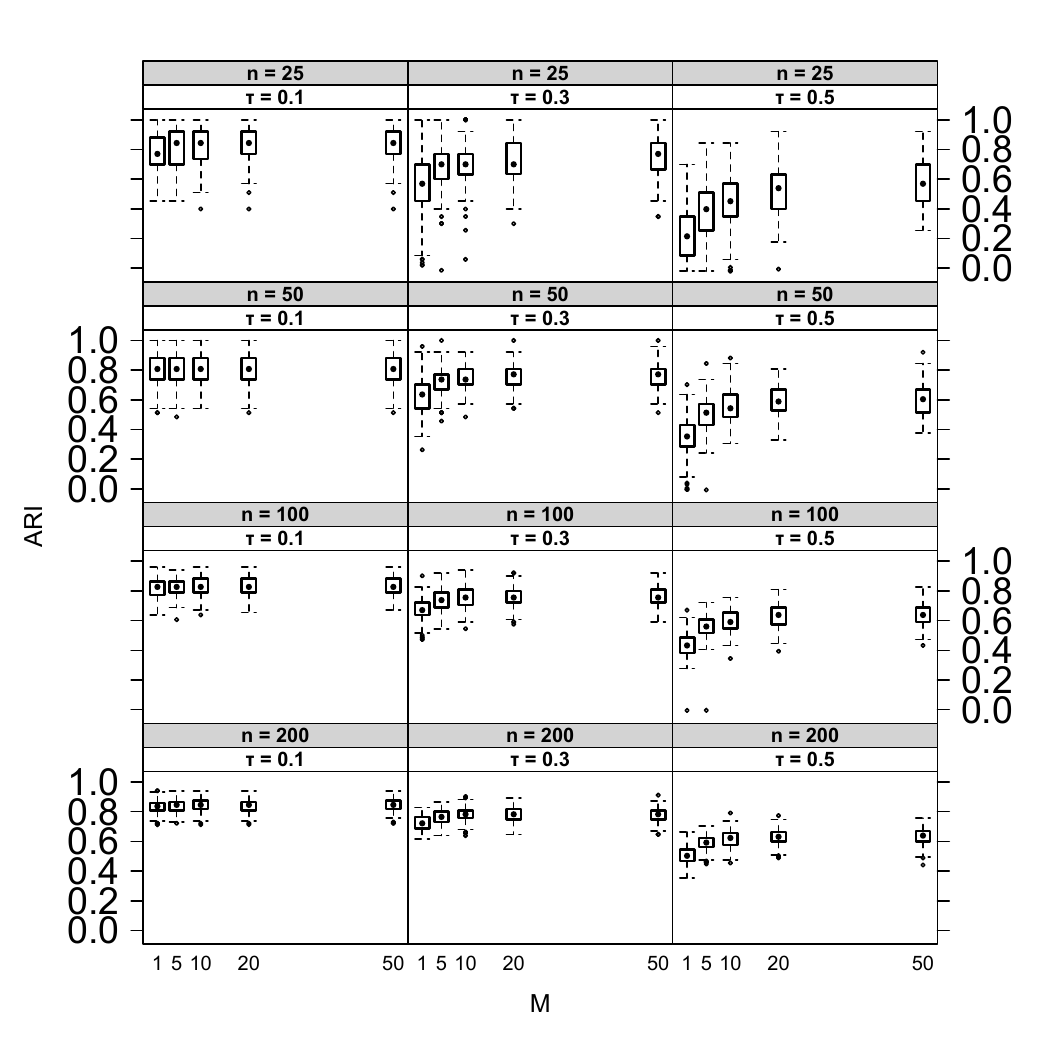}
\caption{ARI distribution for $\rho=0.3$}
\end{subfigure}\begin{subfigure}[t]{.5\textwidth}
\includegraphics[trim={2.2cm .7cm .65cm 1cm},clip,height=10cm,width=7cm]{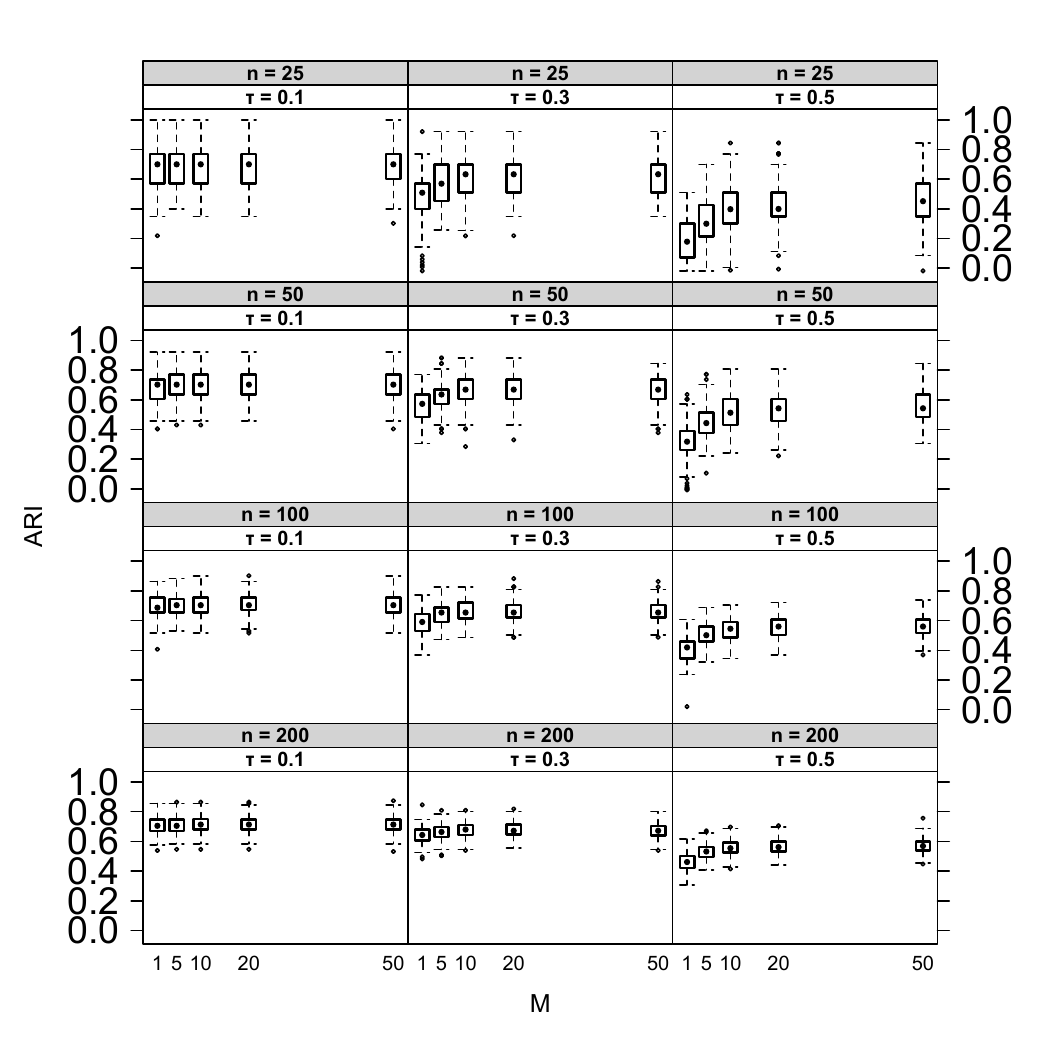}
\caption{ARI distribution for $\rho=0.6$}
\end{subfigure}
\caption{Accuracy of the clustering procedure according to $\Nbtab$: adjusted rand index over the $\Nbsim=200$ generated data sets varying by the number of individuals ($\Nbind$), the correlation between variables ($\rho$) and the proportion of missing values ($\tau$) generated under a MCAR mechanism. Data sets are imputed by JM-DP varying by the number of imputed data sets ($\Nbtab$). For each data set, clustering is performed using k-means clustering and consensus clustering is performed using NMF.\label{figM_ari}}
\end{center}
\end{figure}

As underlined in Section \ref{sectioninst}, the instability given by the second rule does not depend (in expectation) on the number of imputed data sets (if $\Nbtab\geq 2$). As regard the partition accuracy, a large number of imputed data sets should bring the consensus partition closer to the partition obtained from full data. Figure \ref{figM_ari} reports the influence of $\Nbtab$ on the ARI for MI using JM-DP and NMF consensus.

For all configurations, a large value of $\Nbtab$ tends to increase the ARI meaning a large number of imputed data sets tends to increase clustering accuracy. The increase is even more important as the proportion of missing values is large or as the number of individuals is small.

Similar results have been observed when data are imputed by FCS-RF (Figure \ref{figM_ari_mcarrf} in Appendix) or when a MAR mechanism is considered (see Figure \ref{figM_ari_marrf} in Appendix for imputation by FCS-RF and Figure \ref{figM_ari_marmixture} for imputation by JM-DP).\\

\begin{figure}
\begin{center}
\begin{subfigure}[t]{.5\textwidth}
\includegraphics[trim={.4cm .7cm 1.6cm 1cm},scale=.45,clip,height=10cm,width=7cm]{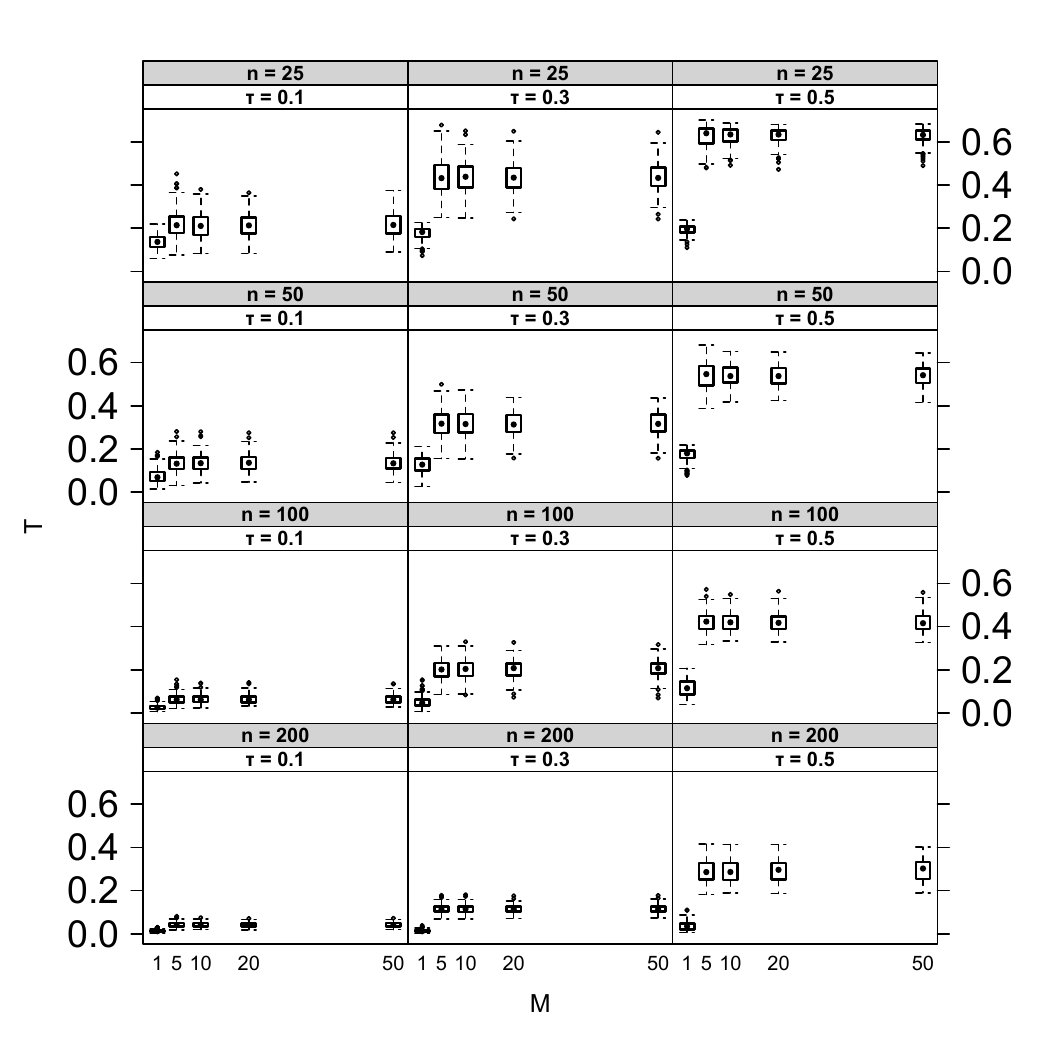}
\caption{$T$ distribution for $\rho=0.3$}
\end{subfigure}\begin{subfigure}[t]{.5\textwidth}
\includegraphics[trim={2.2cm .7cm .65cm 1cm},clip,height=10cm,width=7cm]{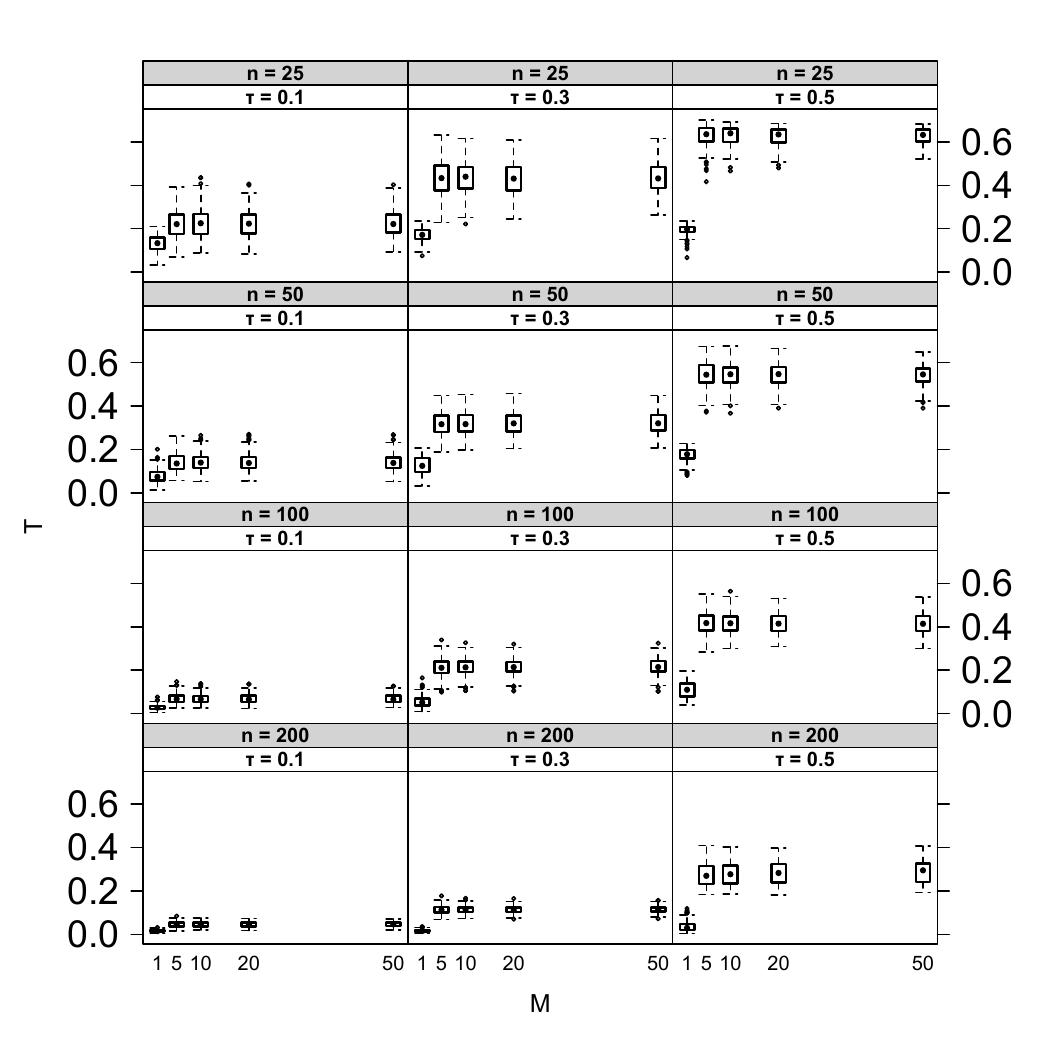}
\caption{$T$ distribution for $\rho=0.6$}
\end{subfigure}
\caption{Instability according to $\Nbtab$: total instability ($T$) over the $\Nbsim=200$ generated data sets varying by the number of individuals ($\Nbind$), the correlation between variables ($\rho$) and the proportion of missing values ($\tau$) generated under a MCAR mechanism. Data sets are imputed by JM-DP varying by the number of imputed data sets ($\Nbtab$). For each data set, clustering is performed using k-means clustering and consensus clustering is performed using NMF.\label{figM_T}}
\end{center}
\end{figure}

Figure \ref{figM_T} reports the influence of $\Nbtab$ on the instability for MI using JM-DP and NMF consensus.

As expected, for $\Nbtab\geq2$ the instability is constant whatever the proportion of missing values, the number of individuals or the correlation between variables. Similar results are observed when data are generated under a MAR mechanism or when data are imputed by FCS-RF (see Figures \ref{figM_T_mixturemar}, \ref{figM_T_rfmcar}, \ref{figM_T_rfmar} in Appendix).

\subsection{Complement: number of clusters}
As written in the introduction, having an instability measure with missing values can provide a way for estimating the number of clusters from incomplete data. To assess the second rule with regards to this goal, a complementary simulation study is conducted by considering a grid for the number of clusters. More precisely, after multiple imputation, k-means clustering is applied for $\Nbgroup$ in $\lbrace 2, \dots, 5\rbrace$. By applying the second rule, a value of instability $T_\Nbgroup$ is obtained from each value from the grid. The estimated number of clusters is given by
\begin{equation}
    argmin_{\Nbgroup\in \{1,...,\Nbgroup_{max}\}}\ T_\Nbgroup.
\end{equation}
Results are reported in Table \ref{tablenbclust}.
\begin{table}[h!]\small
\centering
\caption{Estimated numbers of clusters based on the second rule: frequencies of estimated values within a grid between 2 and 5 clusters over the $\Nbsim=200$ generated data sets for various number of individuals ($\Nbind$), correlation between variables ($\rho$), missing data mechanisms (MCAR and MAR) and proportion of missing values ($\tau$). The expected number of clusters is $\Nbgroup=2$. Two imputation methods are investigated (JM-DP or FCS-RF) using $\Nbtab=50$ imputed data sets. For each case, clustering is performed by k-means. As benchmark, estimated numbers obtained by applying k-means clustering on full data (Full) or complete-case analysis (CCA) are also reported (not possible for a large proportion of missing values). \label{tablenbclust}}
\begin{tabular}{rrrrrrrrrrrrrrrrrrrr}
  \hline
   \multicolumn{1}{c}{} & \multicolumn{1}{c}{} & \multicolumn{1}{c}{} & \multicolumn{1}{c}{} & \multicolumn{4}{c}{JM-DP} & \multicolumn{4}{c}{FCS-RF} & \multicolumn{4}{c}{CCA} & \multicolumn{4}{c}{Full} \\ \hline  $\Nbind$ & $\rho$ & mech &$\tau$ &  2&3&4&5&  2&3&4&5&  2&3&4&5&  2&3&4&5\\
  \hline
50 & 0.3 & MCAR & 0.1 & 39 & 161 & 0 & 0 & 0 & 4 & 46 & 150 & 52 & 4 & 13 & 131 & 192 & 0 & 0 & 8 \\ 
  50 & 0.3 & MCAR & 0.3 & 4 & 196 & 0 & 0 & 0 & 0 & 2 & 198 & 0 & 0 & 0 & 0 & 195 & 1 & 0 & 4 \\ 
  50 & 0.3 & MCAR & 0.5 & 0 & 178 & 0 & 22 & 0 & 0 & 5 & 195 & 0 & 0 & 0 & 0 & 195 & 1 & 0 & 4 \\ 
  50 & 0.3 & MAR & 0.1 & 29 & 171 & 0 & 0 & 0 & 2 & 56 & 142 & 156 & 1 & 2 & 41 & 192 & 1 & 0 & 7 \\ 
  50 & 0.3 & MAR & 0.3 & 0 & 200 & 0 & 0 & 0 & 0 & 50 & 150 & 37 & 12 & 10 & 139 & 194 & 0 & 0 & 6 \\ 
  50 & 0.3 & MAR & 0.5 & 0 & 166 & 1 & 33 & 0 & 0 & 61 & 139 & 1 & 0 & 0 & 3 & 195 & 1 & 0 & 4 \\ 
  50 & 0.6 & MCAR & 0.1 & 28 & 172 & 0 & 0 & 0 & 1 & 50 & 149 & 57 & 3 & 10 & 130 & 187 & 0 & 0 & 13 \\ 
  50 & 0.6 & MCAR & 0.3 & 1 & 199 & 0 & 0 & 0 & 0 & 2 & 198 & 0 & 0 & 0 & 0 & 192 & 0 & 0 & 8 \\ 
  50 & 0.6 & MCAR & 0.5 & 0 & 182 & 0 & 18 & 0 & 0 & 1 & 199 & 0 & 0 & 0 & 0 & 192 & 0 & 0 & 8 \\ 
  50 & 0.6 & MAR & 0.1 & 14 & 186 & 0 & 0 & 2 & 1 & 63 & 134 & 149 & 2 & 1 & 48 & 190 & 0 & 0 & 10 \\ 
  50 & 0.6 & MAR & 0.3 & 1 & 199 & 0 & 0 & 0 & 0 & 29 & 171 & 39 & 2 & 14 & 143 & 190 & 0 & 0 & 10 \\ 
  50 & 0.6 & MAR & 0.5 & 0 & 179 & 0 & 21 & 0 & 0 & 54 & 146 & 0 & 0 & 0 & 8 & 192 & 0 & 0 & 8 \\ 
  200 & 0.3 & MCAR & 0.1 & 200 & 0 & 0 & 0 & 2 & 1 & 148 & 49 & 198 & 0 & 0 & 2 & 200 & 0 & 0 & 0 \\ 
  200 & 0.3 & MCAR & 0.3 & 197 & 3 & 0 & 0 & 0 & 0 & 193 & 7 & 1 & 1 & 3 & 24 & 200 & 0 & 0 & 0 \\ 
  200 & 0.3 & MCAR & 0.5 & 94 & 106 & 0 & 0 & 0 & 0 & 101 & 99 & 0 & 0 & 0 & 0 & 200 & 0 & 0 & 0 \\ 
  200 & 0.3 & MAR & 0.1 & 199 & 1 & 0 & 0 & 2 & 2 & 133 & 63 & 200 & 0 & 0 & 0 & 200 & 0 & 0 & 0 \\ 
  200 & 0.3 & MAR & 0.3 & 195 & 5 & 0 & 0 & 0 & 0 & 173 & 27 & 193 & 0 & 1 & 6 & 200 & 0 & 0 & 0 \\ 
  200 & 0.3 & MAR & 0.5 & 120 & 80 & 0 & 0 & 0 & 0 & 172 & 28 & 96 & 1 & 9 & 94 & 200 & 0 & 0 & 0 \\ 
  200 & 0.6 & MCAR & 0.1 & 200 & 0 & 0 & 0 & 2 & 2 & 147 & 49 & 196 & 0 & 0 & 4 & 200 & 0 & 0 & 0 \\ 
  200 & 0.6 & MCAR & 0.3 & 200 & 0 & 0 & 0 & 0 & 0 & 183 & 17 & 1 & 1 & 0 & 15 & 200 & 0 & 0 & 0 \\ 
  200 & 0.6 & MCAR & 0.5 & 90 & 110 & 0 & 0 & 0 & 0 & 86 & 114 & 0 & 0 & 0 & 0 & 200 & 0 & 0 & 0 \\ 
  200 & 0.6 & MAR & 0.1 & 199 & 1 & 0 & 0 & 3 & 1 & 125 & 71 & 200 & 0 & 0 & 0 & 200 & 0 & 0 & 0 \\ 
  200 & 0.6 & MAR & 0.3 & 200 & 0 & 0 & 0 & 0 & 0 & 151 & 49 & 186 & 0 & 0 & 14 & 200 & 0 & 0 & 0 \\ 
  200 & 0.6 & MAR & 0.5 & 139 & 61 & 0 & 0 & 0 & 0 & 151 & 49 & 87 & 1 & 4 & 108 & 200 & 0 & 0 & 0 \\
   \hline
\end{tabular}
\end{table}
The number of clusters is accurately estimated when performing imputation by JM-DP, but a small number of observations or a large proportion of missing values leads to an upper bias. This behaviour is similar to complete case analysis (even if the method cannot always be applied). Results when using FCS-RF leads to $\Nbgroup=4$ or $\Nbgroup=5$ in most of the cases. Better performances of JM-DP compared to FCS-RF were expected since JM-DP is well tailored to account for the clustered structure of observations \citep{Audigier21}.
\section{Application}
In addition to estimating the instability in clustering, the second rule can be used for tuning the number of clusters with missing values. The \textit{animals} data set from the $cluster$ R package \citep{clusterpkg} is used as an example \citep{Kaufman90}. It describes 20 animals by six binary variables (warm-blooded, can fly, vertebrate, endangered, live in groups, have hair). Five individuals are incomplete (see Table \ref{animals} in Appendix).

\begin{figure}
\begin{center}
\includegraphics[scale=.37]{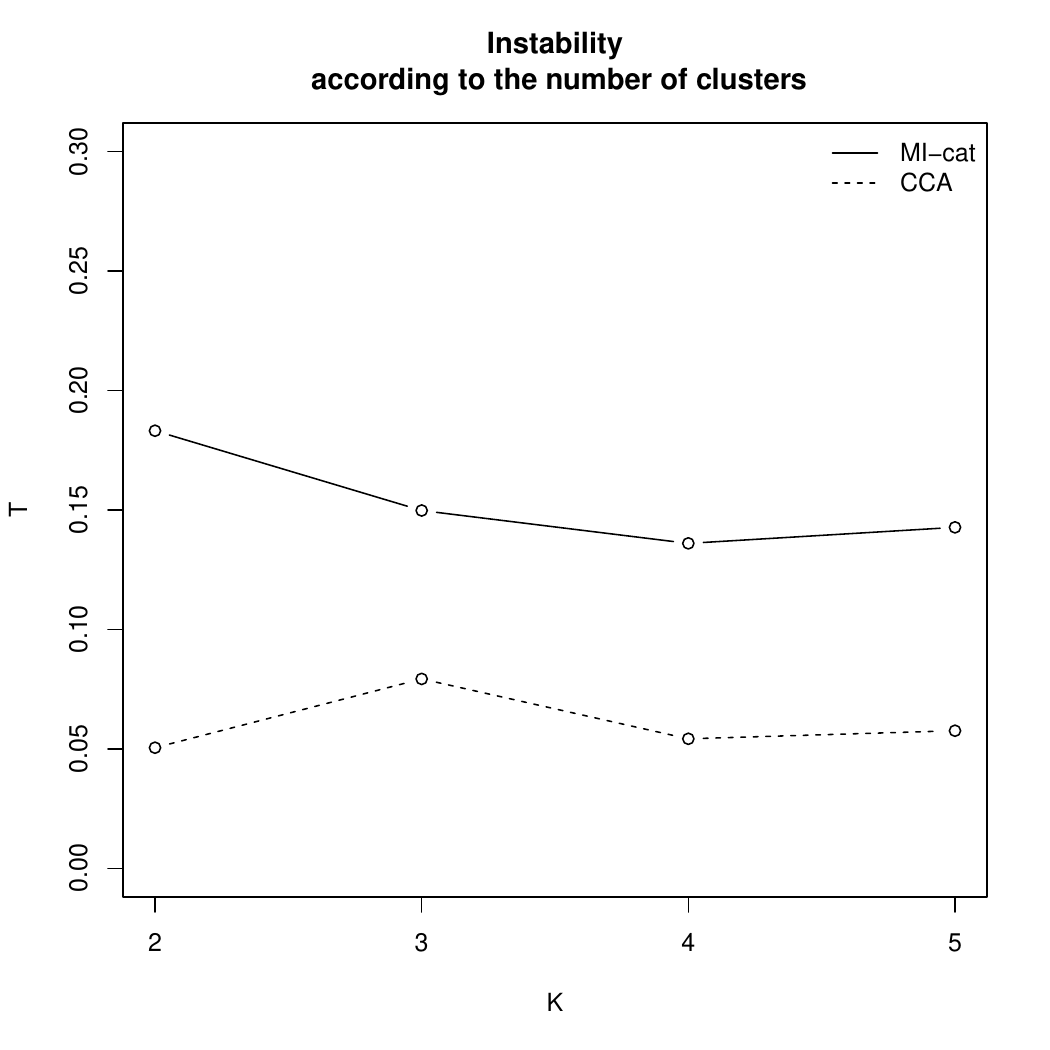}
\caption{Instability estimation in hierarchical clustering according to the number of clusters. Data are imputed using a log-linear model (MI-cat) with $\Nbtab=50$ imputed data sets. The instability with complete-case analysis (CCA) is also reported.\label{animalsk}}
\end{center}
\end{figure}

We propose to perform hierarchical clustering using the flexible UPGMA method \citep{Belbin92}. Flexible UPGMA can be seen as a generalization of the average method which ensures the desirable monotonicity property of the algorithm. For achieving this goal, data are imputed $\Nbtab=50$ times according to a log-linear model \citep{Schafer97}, which is considered as the gold standard for binary variables. Then, hierarchical clustering is applied on each imputed data set for a given number of clusters $\Nbgroup$. Finally, the clustering instability $T$ is assessed using the second rule. This process is repeated for $\Nbgroup$ varying in $\{2,3,4,5\}$. Results are presented in Figure \ref{animalsk} and instability using only complete cases is also reported.

Using MI, the instability is the smallest for $\Nbgroup=4$ clusters, suggesting a consensus clustering in 4 clusters. Results for complete-case analysis are less clear, but a partition in two clusters could be suggested. A larger number could be inappropriate compared to the number of complete cases (15).

\begin{figure}
\begin{center}
\begin{subfigure}[t]{.45\textwidth}
\includegraphics[scale=.37]{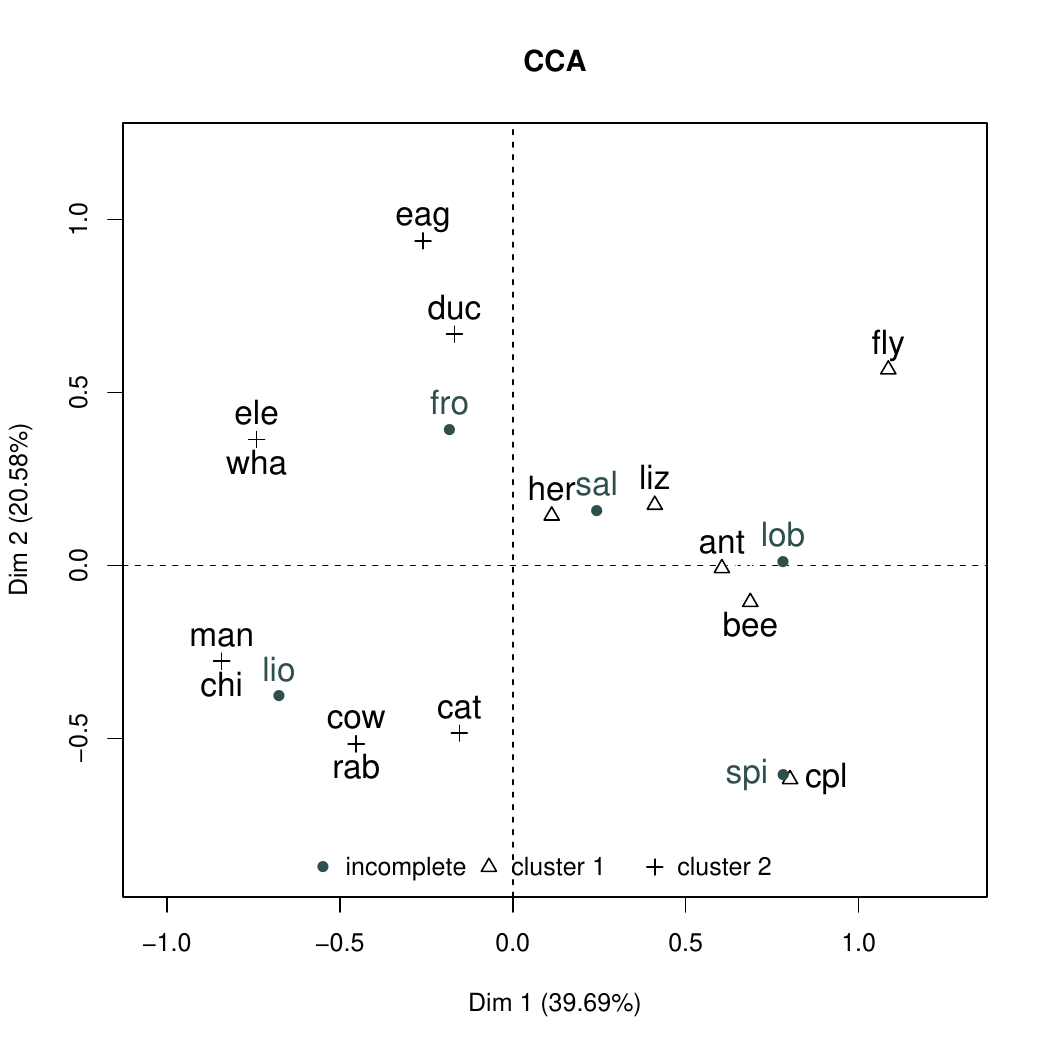}
\caption{Hierarchical clustering in two clusters using complete-case analysis. Incomplete individuals are colored in gray.}
\end{subfigure}\begin{subfigure}[t]{.47\textwidth}
\includegraphics[scale=.37]{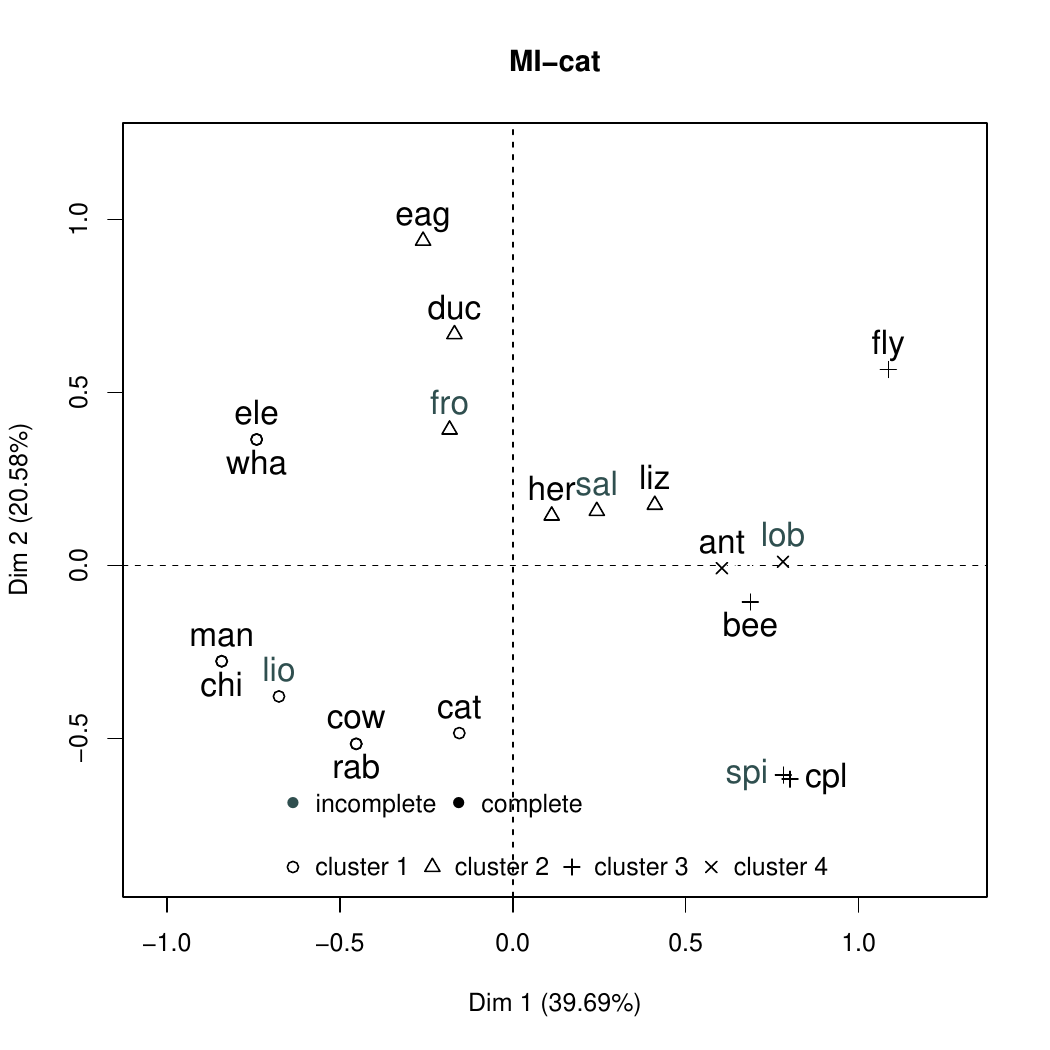}
\caption{Hierarchical clustering in four clusters using consensus clustering (NMF) after MI under log-linear model ($\Nbtab=50$).}
\end{subfigure}
\caption{Animals data set: visualization of partitions through the principal factor map obtained using the iterative MCA algorithm \citep{Josse12}. \label{mca}}
\end{center}
\end{figure}

Comparison between the consensus partition obtained by NMF in four clusters and the one obtained by hierarchical clustering on complete cases in two clusters are presented through a principal factor map (Figure \ref{mca}).

The partition obtained by complete-case analysis gathers mammals with birds in the first cluster, while insects and fishes are gathers in the second. On the opposite, consensus clustering after MI in four clusters isolates mammals in the first cluster, or insects in the fourth. Furthermore, it allows suitable clustering of incomplete individuals: lion among mammals, spider among insects, salmon and frog with herring, lobster with ant (both have the same observed profile \textit{cf} Table \ref{animals} in Appendix).

\section{Discussion}
Multiple imputation is a widely used method for dealing with missing values. However, applying Rubin's rules with clustering remained unclear: 1) how to pool the partitions obtained from each imputed data set?  In this paper, we argue to use median partition-based methods for pooling partitions. In particular, NMF methods are theoretically and computationally attractive for achieving this goal. 2) How to assess the instability of the clustering with missing values? Based on \citet{Fang12}, we propose a new rule for assessing the stability with missing values. 
An associated R package entitled \textit{clusterMI} is available at the web page of the first author.

From a practical point of view, the first rule provides accurate clustering with missing values. Indeed, even without missing data, \citet{Dudoit03} have shown bagging procedures based on bootstrap improve clustering accuracy. By generating $\Nbtab$ times the imputed values and aggregating the partitions obtained from each imputed data set, a similar improvement is observed with multiple imputation. It has been highlighted that the accuracy is sensitive to $\Nbtab$, particularly when the number of individuals is small or the proportion of missing values is large. Simulations show $\Nbtab=50$ is generally enough, but $\Nbtab$ can be tuned by investigating the evolution of the pooled partition according to the number of imputed data sets.

The second rule allows calculation of an additional between instability $B$ related to missing values. This instability has the advantage to be robust to the choice of $\Nbtab$. Availability of a between instability is precious in practice for several uses. Firstly, it provides a new way for dealing with the number of clusters when data are incomplete. This is particularly useful for distance-based clustering methods like k-means or k-medoids. Secondly, the ratio $B/T$ provides a new way to highlight how the partition is robust to the missing values \citep{VB18}.\\

In this work, we assumed data were already imputed. It could be also interesting to investigate more deeply the suitable imputation method according to the clustering method applied on each imputed data set. The topic is commonly discussed under the term \textit{congeniality} \citep{Meng94,Schafer03}. Simulation results subtend a sensitivity to the imputation method both in terms of accuracy and instability which is substantiated by recent research \citep{Audigier21}. We also assumed data were missing at random, but beyond the difficulty to impute non-missing at random data, the proposed rules could be directly applied.

Furthermore, we assumed all contributory partitions have the same number of clusters $\Nbgroup$, but the methodology can be directly applied for various number of clusters, like in hierarchical clustering where the number is generally unknown in advance. Indeed, NMF consensus clustering is essentially based on the average of the connectivity matrices associated to each contributory partition. Such an average can be obtained whatever the number of clusters since the dimensions of each connectivity matrix depends only on the number of individuals which is constant for all partitions. The only requirement is that the clustering method allows classification for new individuals. Even if this classification is always possible, certain classification methods are connected to certain clustering methods. For instance, classification using the closest centroid is suitable for k-means or k-medoids, while quadratic discriminant analysis could be more reliable for Gaussian mixtures. For other clustering methods, anyone of these classification methods could be used, but the robustness to the stability measurement should require more research.

Finally, several NMF-based methods are available for partitions pooling. In this paper, we focus on the multiplicative rules method as proposed in \citet{Li07} which is the most common method, but which is not necessarily the most efficient. Among alternatives, alternating least squares algorithms are notably recommended for large scale data \citep{qn_nmf}.

\bibliographystyle{spbasic}      
\bibliography{references}   

\appendix

\section*{Appendix}

\subsection*{Partitions pooling}
\begin{sidewaystable}
\centering
\caption{Accuracy of the clustering procedure under a MCAR mechanism: average adjusted rand index (mean) and interquartile range (IQ) over the $\Nbsim=200$ generated data sets varying by the number of individuals ($\Nbind$), the correlation between variables ($\rho$) and the proportion of missing values ($\tau$) for various imputation methods (JM-DP or FCS-RF), various consensus methods (NMF or SAOM). For each case, clustering is performed using k-means clustering. As benchmark, ARI obtained by applying k-means on complete-cases (CCA), using k-pod algorithm (kpod), on full data (Full) or using a bagging procedure (Full-boot) are also reported. \label{tablearimcar}} 
\begin{tabular}{lllrrrrrrrrrrrrrrrr}
  \hline
     \multicolumn{1}{c}{} & \multicolumn{1}{c}{} & \multicolumn{1}{c}{} & \multicolumn{4}{c}{NMF} & \multicolumn{4}{c}{SAOM} &  \\
    \multicolumn{1}{c}{} & \multicolumn{1}{c}{} & \multicolumn{1}{c}{} & \multicolumn{2}{c}{JM-DP} & \multicolumn{2}{c}{FCS-RF} & \multicolumn{2}{c}{JM-DP} & \multicolumn{2}{c}{FCS-RF} & \multicolumn{2}{c}{kpod} & \multicolumn{2}{c}{CCA} & \multicolumn{2}{c}{Full} & \multicolumn{2}{c}{Full-boot} \\
   $\Nbind$ & $\rho$ & $\tau$ & mean  & IQ  & mean & IQ & mean & IQ & mean  & IQ & mean & IQ & mean & IQ & mean & IQ & mean & IQ \\ 
  \hline
25 & 0.3 & 0.1 & 0.82 & 0.15 & 0.80 & 0.22 & 0.82 & 0.15 & 0.80 & 0.22 & 0.83 & 0.15 & 0.37 & 0.25 & 0.83 & 0.15 & 0.83 & 0.15 \\ 
  25 & 0.3 & 0.3 & 0.75 & 0.16 & 0.73 & 0.21 & 0.73 & 0.21 & 0.74 & 0.21 & 0.75 & 0.14 & 0.03 & 0.04 & 0.83 & 0.15 & 0.83 & 0.15 \\ 
  25 & 0.3 & 0.5 & 0.56 & 0.25 & 0.58 & 0.25 & 0.57 & 0.25 & 0.57 & 0.25 & 0.52 & 0.30 & 0.00 & 0.02 & 0.83 & 0.15 & 0.83 & 0.15 \\ 
  25 & 0.6 & 0.1 & 0.70 & 0.15 & 0.69 & 0.20 & 0.70 & 0.20 & 0.68 & 0.20 & 0.73 & 0.21 & 0.32 & 0.23 & 0.71 & 0.21 & 0.71 & 0.21 \\ 
  25 & 0.6 & 0.3 & 0.64 & 0.19 & 0.61 & 0.19 & 0.63 & 0.19 & 0.60 & 0.19 & 0.64 & 0.20 & 0.04 & 0.05 & 0.71 & 0.21 & 0.71 & 0.21 \\ 
  25 & 0.6 & 0.5 & 0.46 & 0.22 & 0.47 & 0.17 & 0.46 & 0.20 & 0.47 & 0.20 & 0.43 & 0.27 & -0.00 & 0.02 & 0.71 & 0.21 & 0.71 & 0.21 \\ 
  50 & 0.3 & 0.1 & 0.81 & 0.15 & 0.80 & 0.15 & 0.81 & 0.15 & 0.80 & 0.15 & 0.83 & 0.11 & 0.38 & 0.17 & 0.84 & 0.15 & 0.83 & 0.11 \\ 
  50 & 0.3 & 0.3 & 0.76 & 0.11 & 0.74 & 0.14 & 0.75 & 0.11 & 0.73 & 0.12 & 0.78 & 0.14 & 0.13 & 0.20 & 0.84 & 0.15 & 0.83 & 0.11 \\ 
  50 & 0.3 & 0.5 & 0.61 & 0.16 & 0.60 & 0.16 & 0.60 & 0.13 & 0.59 & 0.16 & 0.59 & 0.16 & -0.00 & 0.01 & 0.84 & 0.15 & 0.83 & 0.11 \\ 
  50 & 0.6 & 0.1 & 0.70 & 0.14 & 0.68 & 0.13 & 0.70 & 0.14 & 0.68 & 0.13 & 0.72 & 0.17 & 0.34 & 0.14 & 0.71 & 0.14 & 0.71 & 0.14 \\ 
  50 & 0.6 & 0.3 & 0.66 & 0.13 & 0.62 & 0.10 & 0.65 & 0.12 & 0.61 & 0.13 & 0.68 & 0.17 & 0.11 & 0.16 & 0.71 & 0.14 & 0.71 & 0.14 \\ 
  50 & 0.6 & 0.5 & 0.55 & 0.15 & 0.51 & 0.12 & 0.54 & 0.15 & 0.50 & 0.14 & 0.53 & 0.15 & -0.00 & 0.01 & 0.71 & 0.14 & 0.71 & 0.14 \\ 
  100 & 0.3 & 0.1 & 0.83 & 0.09 & 0.81 & 0.07 & 0.83 & 0.08 & 0.81 & 0.08 & 0.84 & 0.09 & 0.39 & 0.12 & 0.84 & 0.07 & 0.84 & 0.07 \\ 
  100 & 0.3 & 0.3 & 0.76 & 0.09 & 0.73 & 0.09 & 0.75 & 0.08 & 0.72 & 0.08 & 0.77 & 0.09 & 0.21 & 0.28 & 0.84 & 0.07 & 0.84 & 0.07 \\ 
  100 & 0.3 & 0.5 & 0.64 & 0.10 & 0.61 & 0.08 & 0.62 & 0.10 & 0.59 & 0.09 & 0.60 & 0.09 & 0.00 & 0.01 & 0.84 & 0.08 & 0.84 & 0.07 \\ 
  100 & 0.6 & 0.1 & 0.70 & 0.10 & 0.68 & 0.08 & 0.70 & 0.09 & 0.68 & 0.09 & 0.72 & 0.10 & 0.33 & 0.11 & 0.70 & 0.10 & 0.70 & 0.10 \\ 
  100 & 0.6 & 0.3 & 0.66 & 0.08 & 0.60 & 0.10 & 0.65 & 0.10 & 0.59 & 0.10 & 0.68 & 0.08 & 0.20 & 0.25 & 0.70 & 0.10 & 0.70 & 0.10 \\ 
  100 & 0.6 & 0.5 & 0.57 & 0.09 & 0.49 & 0.09 & 0.55 & 0.09 & 0.48 & 0.08 & 0.55 & 0.10 & 0.00 & 0.01 & 0.70 & 0.10 & 0.70 & 0.10 \\ 
  200 & 0.3 & 0.1 & 0.84 & 0.06 & 0.82 & 0.05 & 0.84 & 0.05 & 0.81 & 0.06 & 0.85 & 0.06 & 0.39 & 0.08 & 0.84 & 0.05 & 0.84 & 0.06 \\ 
  200 & 0.3 & 0.3 & 0.78 & 0.06 & 0.74 & 0.06 & 0.77 & 0.06 & 0.72 & 0.07 & 0.78 & 0.06 & 0.21 & 0.17 & 0.84 & 0.05 & 0.84 & 0.06 \\ 
  200 & 0.3 & 0.5 & 0.63 & 0.07 & 0.60 & 0.07 & 0.61 & 0.06 & 0.58 & 0.07 & 0.61 & 0.06 & -0.00 & 0.00 & 0.84 & 0.05 & 0.84 & 0.06 \\ 
  200 & 0.6 & 0.1 & 0.71 & 0.07 & 0.69 & 0.07 & 0.71 & 0.07 & 0.68 & 0.08 & 0.72 & 0.08 & 0.34 & 0.07 & 0.71 & 0.08 & 0.71 & 0.08 \\ 
  200 & 0.6 & 0.3 & 0.67 & 0.07 & 0.61 & 0.07 & 0.67 & 0.07 & 0.60 & 0.08 & 0.68 & 0.07 & 0.20 & 0.13 & 0.71 & 0.08 & 0.71 & 0.08 \\ 
  200 & 0.6 & 0.5 & 0.57 & 0.06 & 0.50 & 0.08 & 0.55 & 0.07 & 0.48 & 0.07 & 0.56 & 0.08 & 0.00 & 0.00 & 0.71 & 0.08 & 0.71 & 0.08 \\ 
   \hline
  \end{tabular}
\end{sidewaystable}

\begin{sidewaystable}
\centering
\caption{Accuracy of the clustering procedure under a MAR mechanism: average adjusted rand index (mean) and interquartile range (IQ) over the $\Nbsim=200$ generated data sets varying by the number of individuals ($\Nbind$), the correlation between variables ($\rho$) and the proportion of missing values ($\tau$) for various imputation methods (JM-DP or FCS-RF), various consensus methods (NMF or SAOM). For each case, clustering is performed using k-means clustering. As benchmark, ARI obtained by applying k-means on complete-cases (CCA), using k-pod algorithm (kpod), on full data (Full) or using a bagging procedure (Full-boot) are also reported. \label{tablearimar}} 
\begin{tabular}{lllrrrrrrrrrrrrrrrr}
  \hline
     \multicolumn{1}{c}{} & \multicolumn{1}{c}{} & \multicolumn{1}{c}{} & \multicolumn{4}{c}{NMF} & \multicolumn{4}{c}{SAOM} &  \\
    \multicolumn{1}{c}{} & \multicolumn{1}{c}{} & \multicolumn{1}{c}{} & \multicolumn{2}{c}{JM-DP} & \multicolumn{2}{c}{FCS-RF} & \multicolumn{2}{c}{JM-DP} & \multicolumn{2}{c}{FCS-RF} & \multicolumn{2}{c}{kpod} & \multicolumn{2}{c}{CCA} & \multicolumn{2}{c}{Full} & \multicolumn{2}{c}{Full-boot} \\
   $\Nbind$ & $\rho$ & $\tau$ & mean  & IQ  & mean & IQ & mean & IQ & mean  & IQ & mean & IQ & mean & IQ & mean & IQ & mean & IQ \\ 
  \hline
 25 & 0.3 & 0.1 & 0.79 & 0.14 & 0.79 & 0.22 & 0.79 & 0.14 & 0.79 & 0.22 & 0.81 & 0.15 & 0.54 & 0.18 & 0.83 & 0.15 & 0.83 & 0.15 \\ 
  25 & 0.3 & 0.3 & 0.64 & 0.26 & 0.64 & 0.19 & 0.64 & 0.26 & 0.64 & 0.16 & 0.56 & 0.25 & 0.31 & 0.22 & 0.83 & 0.15 & 0.83 & 0.15 \\ 
  25 & 0.3 & 0.5 & 0.42 & 0.16 & 0.43 & 0.16 & 0.42 & 0.19 & 0.44 & 0.16 & 0.31 & 0.18 & 0.15 & 0.19 & 0.83 & 0.15 & 0.83 & 0.15 \\ 
  25 & 0.6 & 0.1 & 0.68 & 0.20 & 0.67 & 0.20 & 0.68 & 0.20 & 0.66 & 0.20 & 0.70 & 0.27 & 0.47 & 0.22 & 0.71 & 0.21 & 0.71 & 0.21 \\ 
  25 & 0.6 & 0.3 & 0.58 & 0.25 & 0.55 & 0.18 & 0.57 & 0.21 & 0.55 & 0.18 & 0.53 & 0.19 & 0.27 & 0.17 & 0.71 & 0.21 & 0.71 & 0.21 \\ 
  25 & 0.6 & 0.5 & 0.38 & 0.15 & 0.38 & 0.19 & 0.39 & 0.15 & 0.37 & 0.15 & 0.28 & 0.22 & 0.12 & 0.19 & 0.71 & 0.21 & 0.71 & 0.21 \\ 
  50 & 0.3 & 0.1 & 0.80 & 0.15 & 0.79 & 0.11 & 0.80 & 0.13 & 0.79 & 0.11 & 0.80 & 0.15 & 0.54 & 0.18 & 0.84 & 0.15 & 0.83 & 0.11 \\ 
  50 & 0.3 & 0.3 & 0.67 & 0.13 & 0.64 & 0.13 & 0.67 & 0.13 & 0.64 & 0.13 & 0.59 & 0.16 & 0.29 & 0.13 & 0.84 & 0.15 & 0.83 & 0.11 \\ 
  50 & 0.3 & 0.5 & 0.47 & 0.12 & 0.46 & 0.11 & 0.46 & 0.12 & 0.45 & 0.13 & 0.31 & 0.14 & 0.17 & 0.15 & 0.84 & 0.15 & 0.83 & 0.11 \\ 
  50 & 0.6 & 0.1 & 0.68 & 0.17 & 0.66 & 0.13 & 0.68 & 0.17 & 0.66 & 0.13 & 0.69 & 0.14 & 0.46 & 0.13 & 0.71 & 0.14 & 0.71 & 0.14 \\ 
  50 & 0.6 & 0.3 & 0.59 & 0.16 & 0.56 & 0.15 & 0.58 & 0.14 & 0.56 & 0.15 & 0.54 & 0.15 & 0.28 & 0.15 & 0.71 & 0.14 & 0.71 & 0.14 \\ 
  50 & 0.6 & 0.5 & 0.40 & 0.13 & 0.38 & 0.15 & 0.40 & 0.13 & 0.38 & 0.14 & 0.28 & 0.11 & 0.15 & 0.12 & 0.71 & 0.14 & 0.71 & 0.14 \\ 
  100 & 0.3 & 0.1 & 0.81 & 0.07 & 0.79 & 0.08 & 0.81 & 0.07 & 0.79 & 0.09 & 0.80 & 0.07 & 0.55 & 0.11 & 0.84 & 0.08 & 0.84 & 0.07 \\ 
  100 & 0.3 & 0.3 & 0.67 & 0.10 & 0.65 & 0.09 & 0.66 & 0.09 & 0.64 & 0.08 & 0.58 & 0.11 & 0.32 & 0.11 & 0.84 & 0.08 & 0.84 & 0.07 \\ 
  100 & 0.3 & 0.5 & 0.47 & 0.10 & 0.45 & 0.09 & 0.46 & 0.09 & 0.45 & 0.10 & 0.32 & 0.09 & 0.18 & 0.10 & 0.84 & 0.07 & 0.84 & 0.07 \\ 
  100 & 0.6 & 0.1 & 0.68 & 0.09 & 0.66 & 0.10 & 0.68 & 0.10 & 0.66 & 0.10 & 0.70 & 0.09 & 0.46 & 0.10 & 0.70 & 0.10 & 0.70 & 0.10 \\ 
  100 & 0.6 & 0.3 & 0.59 & 0.11 & 0.54 & 0.10 & 0.58 & 0.10 & 0.54 & 0.10 & 0.55 & 0.09 & 0.27 & 0.10 & 0.70 & 0.10 & 0.70 & 0.10 \\ 
  100 & 0.6 & 0.5 & 0.42 & 0.09 & 0.37 & 0.11 & 0.41 & 0.09 & 0.37 & 0.11 & 0.30 & 0.09 & 0.15 & 0.09 & 0.70 & 0.10 & 0.70 & 0.10 \\ 
  200 & 0.3 & 0.1 & 0.82 & 0.06 & 0.80 & 0.07 & 0.82 & 0.06 & 0.80 & 0.06 & 0.81 & 0.06 & 0.55 & 0.07 & 0.84 & 0.05 & 0.84 & 0.06 \\ 
  200 & 0.3 & 0.3 & 0.69 & 0.07 & 0.65 & 0.06 & 0.68 & 0.07 & 0.65 & 0.06 & 0.59 & 0.07 & 0.31 & 0.07 & 0.84 & 0.05 & 0.84 & 0.06 \\ 
  200 & 0.3 & 0.5 & 0.47 & 0.07 & 0.45 & 0.07 & 0.46 & 0.06 & 0.45 & 0.06 & 0.33 & 0.06 & 0.19 & 0.07 & 0.84 & 0.05 & 0.84 & 0.06 \\ 
  200 & 0.6 & 0.1 & 0.69 & 0.07 & 0.67 & 0.07 & 0.69 & 0.06 & 0.67 & 0.06 & 0.70 & 0.06 & 0.47 & 0.06 & 0.71 & 0.08 & 0.71 & 0.08 \\ 
  200 & 0.6 & 0.3 & 0.59 & 0.07 & 0.54 & 0.07 & 0.58 & 0.07 & 0.54 & 0.07 & 0.55 & 0.07 & 0.28 & 0.07 & 0.71 & 0.08 & 0.71 & 0.08 \\ 
  200 & 0.6 & 0.5 & 0.41 & 0.07 & 0.38 & 0.05 & 0.40 & 0.06 & 0.37 & 0.06 & 0.30 & 0.06 & 0.16 & 0.08 & 0.71 & 0.08 & 0.71 & 0.08 \\
   \hline
  \end{tabular}
\end{sidewaystable}
\clearpage
\subsection*{Instability pooling}
\begin{table}[H]
\centering
\caption{Instability of the clustering procedure under a MCAR mechanism: average within-instability ($\bar{U}$) average between-instability ($B$) and average total instability ($T$) over the $\Nbsim=200$ generated data sets for various number of individuals ($\Nbind$), correlation between variables ($\rho$) and proportion of missing values ($\tau$). Two imputation methods are investigated (JM-DP or FCS-RF) using $\Nbtab=1$ or $\Nbtab=50$ imputed data sets. For each case, clustering is performed by k-means. As benchmark, ARI obtained by applying k-means clustering on full data (Full) or complete-case analysis (CCA) are also reported (not possible with a large proportion of missing values). \label{tableinstnpetitmcar}}
\begin{tabular}{llllrrrrrrrr}
  \hline
   \multicolumn{1}{c}{} & \multicolumn{1}{c}{} & \multicolumn{1}{c}{} & \multicolumn{1}{c}{} & \multicolumn{3}{c}{JM-DP} & \multicolumn{3}{c}{FCS-RF} & \multicolumn{1}{c}{Full} & \multicolumn{1}{c}{CCA} \\$\Nbind$ & $\rho$ & $\tau$ & $M$ &  $\bar{U}$ &  $B$ &  $T$ &  $\bar{U}$ &  $B$ &  $T$ & $T$ & $T$ \\ 
  \hline
25 & 0.3 & 0.1 & 1 & 0.14 & 0.00 & 0.14 & 0.12 & 0.00 & 0.12 & 0.11 & 0.16 \\ 
  25 & 0.3 & 0.1 & 50 & 0.13 & 0.08 & 0.22 & 0.12 & 0.07 & 0.19 & 0.11 & 0.16 \\ 
  25 & 0.3 & 0.3 & 1 & 0.18 & 0.00 & 0.18 & 0.15 & 0.00 & 0.15 & 0.11 &  \\ 
  25 & 0.3 & 0.3 & 50 & 0.17 & 0.26 & 0.43 & 0.15 & 0.20 & 0.35 & 0.11 &  \\ 
  25 & 0.3 & 0.5 & 1 & 0.19 & 0.00 & 0.19 & 0.18 & 0.00 & 0.18 & 0.11 &  \\ 
  25 & 0.3 & 0.5 & 50 & 0.19 & 0.43 & 0.63 & 0.18 & 0.37 & 0.55 & 0.11 &  \\ 
  25 & 0.6 & 0.1 & 1 & 0.13 & 0.00 & 0.13 & 0.12 & 0.00 & 0.12 & 0.11 & 0.16 \\ 
  25 & 0.6 & 0.1 & 50 & 0.13 & 0.09 & 0.22 & 0.12 & 0.07 & 0.20 & 0.11 & 0.16 \\ 
  25 & 0.6 & 0.3 & 1 & 0.17 & 0.00 & 0.17 & 0.15 & 0.00 & 0.15 & 0.11 &  \\ 
  25 & 0.6 & 0.3 & 50 & 0.17 & 0.26 & 0.44 & 0.15 & 0.20 & 0.34 & 0.11 &  \\ 
  25 & 0.6 & 0.5 & 1 & 0.19 & 0.00 & 0.19 & 0.17 & 0.00 & 0.17 & 0.11 &  \\ 
  25 & 0.6 & 0.5 & 50 & 0.19 & 0.43 & 0.63 & 0.17 & 0.37 & 0.54 & 0.11 &  \\ 
  100 & 0.3 & 0.1 & 1 & 0.03 & 0.00 & 0.03 & 0.03 & 0.00 & 0.03 & 0.02 & 0.09 \\ 
  100 & 0.3 & 0.1 & 50 & 0.03 & 0.04 & 0.06 & 0.03 & 0.05 & 0.08 & 0.02 & 0.09 \\ 
  100 & 0.3 & 0.3 & 1 & 0.05 & 0.00 & 0.05 & 0.05 & 0.00 & 0.05 & 0.02 &  \\ 
  100 & 0.3 & 0.3 & 50 & 0.05 & 0.15 & 0.20 & 0.04 & 0.15 & 0.19 & 0.02 &  \\ 
  100 & 0.3 & 0.5 & 1 & 0.12 & 0.00 & 0.12 & 0.08 & 0.00 & 0.08 & 0.02 &  \\ 
  100 & 0.3 & 0.5 & 50 & 0.11 & 0.30 & 0.42 & 0.08 & 0.28 & 0.36 & 0.02 &  \\ 
  100 & 0.6 & 0.1 & 1 & 0.03 & 0.00 & 0.03 & 0.03 & 0.00 & 0.03 & 0.03 & 0.10 \\ 
  100 & 0.6 & 0.1 & 50 & 0.03 & 0.04 & 0.07 & 0.03 & 0.05 & 0.08 & 0.03 & 0.10 \\ 
  100 & 0.6 & 0.3 & 1 & 0.05 & 0.00 & 0.05 & 0.04 & 0.00 & 0.04 & 0.03 &  \\ 
  100 & 0.6 & 0.3 & 50 & 0.06 & 0.16 & 0.22 & 0.05 & 0.14 & 0.19 & 0.03 &  \\ 
  100 & 0.6 & 0.5 & 1 & 0.11 & 0.00 & 0.11 & 0.08 & 0.00 & 0.08 & 0.03 &  \\ 
  100 & 0.6 & 0.5 & 50 & 0.11 & 0.31 & 0.41 & 0.08 & 0.26 & 0.34 & 0.03 &  \\ 
   \hline
  \end{tabular}
\end{table}

\begin{sidewaystable}
\caption{Instability of the clustering procedure under a MAR mechanism: average within-instability ($\bar{U}$) average between-instability ($B$) and average total instability ($T$) over the $\Nbsim=200$ generated data sets for various number of individuals ($\Nbind$), correlation between variables ($\rho$) and proportion of missing values ($\tau$). Two imputation methods are investigated (JM-DP or FCS-RF) using $\Nbtab=1$ or $\Nbtab=50$ imputed data sets. For each case, clustering is performed by k-means. As benchmark, ARI obtained by applying k-means clustering on full data (Full) or complete-case analysis (CCA) are also reported (not possible with a large proportion of missing values).\label{tableinstmar}}\footnotesize
    \begin{subtable}[h]{0.5\textwidth}
\centering
\caption{$\Nbind \in \{25, 50\}$}
\begin{tabular}{p{0.02\textwidth}p{0.02\textwidth}p{0.02\textwidth}p{0.02\textwidth}p{0.04\textwidth}p{0.04\textwidth}p{0.04\textwidth}p{0.04\textwidth}p{0.04\textwidth}p{0.04\textwidth}p{0.04\textwidth}p{0.03\textwidth}}
  \hline
   \multicolumn{1}{c}{} & \multicolumn{1}{c}{} & \multicolumn{1}{c}{} & \multicolumn{1}{c}{} & \multicolumn{3}{c}{JM-DP} & \multicolumn{3}{c}{FCS-RF} & \multicolumn{1}{c}{Full} & \multicolumn{1}{c}{CCA} \\$\Nbind$ & $\rho$ & $\tau$ & $M$ &  $\bar{U}$ &  $B$ &  $T$ &  $\bar{U}$ &  $B$ &  $T$ & $T$ & $T$ \\ 
  \hline
25 & 0.3 & 0.1 & 1 & 0.14 & 0.00 & 0.14 & 0.12 & 0.00 & 0.12 & 0.11 & 0.14 \\ 
  25 & 0.3 & 0.1 & 50 & 0.13 & 0.09 & 0.22 & 0.12 & 0.08 & 0.20 & 0.11 & 0.14 \\ 
  25 & 0.3 & 0.3 & 1 & 0.17 & 0.00 & 0.17 & 0.15 & 0.00 & 0.15 & 0.11 & 0.16 \\ 
  25 & 0.3 & 0.3 & 50 & 0.17 & 0.27 & 0.44 & 0.15 & 0.22 & 0.37 & 0.11 & 0.16 \\ 
  25 & 0.3 & 0.5 & 1 & 0.19 & 0.00 & 0.19 & 0.17 & 0.00 & 0.17 & 0.11 &  \\ 
  25 & 0.3 & 0.5 & 50 & 0.19 & 0.41 & 0.60 & 0.17 & 0.36 & 0.53 & 0.11 &  \\ 
  25 & 0.6 & 0.1 & 1 & 0.13 & 0.00 & 0.13 & 0.13 & 0.00 & 0.13 & 0.11 & 0.13 \\ 
  25 & 0.6 & 0.1 & 50 & 0.13 & 0.10 & 0.23 & 0.12 & 0.08 & 0.20 & 0.11 & 0.13 \\ 
  25 & 0.6 & 0.3 & 1 & 0.17 & 0.00 & 0.17 & 0.15 & 0.00 & 0.15 & 0.11 & 0.15 \\ 
  25 & 0.6 & 0.3 & 50 & 0.17 & 0.26 & 0.42 & 0.15 & 0.21 & 0.35 & 0.11 & 0.15 \\ 
  25 & 0.6 & 0.5 & 1 & 0.19 & 0.00 & 0.19 & 0.17 & 0.00 & 0.17 & 0.11 &  \\ 
  25 & 0.6 & 0.5 & 50 & 0.19 & 0.41 & 0.60 & 0.17 & 0.36 & 0.52 & 0.11 &  \\ 
  50 & 0.3 & 0.1 & 1 & 0.08 & 0.00 & 0.08 & 0.07 & 0.00 & 0.07 & 0.06 & 0.09 \\ 
  50 & 0.3 & 0.1 & 50 & 0.08 & 0.07 & 0.15 & 0.07 & 0.07 & 0.14 & 0.06 & 0.09 \\ 
  50 & 0.3 & 0.3 & 1 & 0.12 & 0.00 & 0.12 & 0.10 & 0.00 & 0.10 & 0.06 & 0.14 \\ 
  50 & 0.3 & 0.3 & 50 & 0.12 & 0.22 & 0.34 & 0.10 & 0.19 & 0.30 & 0.06 & 0.14 \\ 
  50 & 0.3 & 0.5 & 1 & 0.17 & 0.00 & 0.17 & 0.14 & 0.00 & 0.14 & 0.06 & 0.15 \\ 
  50 & 0.3 & 0.5 & 50 & 0.17 & 0.36 & 0.53 & 0.14 & 0.33 & 0.47 & 0.06 & 0.15 \\ 
  50 & 0.6 & 0.1 & 1 & 0.08 & 0.00 & 0.08 & 0.07 & 0.00 & 0.07 & 0.06 & 0.09 \\ 
  50 & 0.6 & 0.1 & 50 & 0.08 & 0.08 & 0.15 & 0.07 & 0.06 & 0.14 & 0.06 & 0.09 \\ 
  50 & 0.6 & 0.3 & 1 & 0.12 & 0.00 & 0.12 & 0.10 & 0.00 & 0.10 & 0.06 & 0.14 \\ 
  50 & 0.6 & 0.3 & 50 & 0.12 & 0.22 & 0.34 & 0.10 & 0.19 & 0.29 & 0.06 & 0.14 \\ 
  50 & 0.6 & 0.5 & 1 & 0.16 & 0.00 & 0.16 & 0.14 & 0.00 & 0.14 & 0.06 & 0.16 \\ 
  50 & 0.6 & 0.5 & 50 & 0.16 & 0.36 & 0.53 & 0.14 & 0.32 & 0.46 & 0.06 & 0.16 \\  
   \hline
  \end{tabular}
\end{subtable}
\begin{subtable}[H]{0.4\textwidth}
\centering
\caption{$\Nbind \in \{100, 200\}$}
\begin{tabular}{p{0.02\textwidth}p{0.02\textwidth}p{0.02\textwidth}p{0.02\textwidth}p{0.04\textwidth}p{0.04\textwidth}p{0.04\textwidth}p{0.04\textwidth}p{0.04\textwidth}p{0.04\textwidth}p{0.04\textwidth}p{0.03\textwidth}}
  \hline
   \multicolumn{1}{c}{} & \multicolumn{1}{c}{} & \multicolumn{1}{c}{} & \multicolumn{1}{c}{} & \multicolumn{3}{c}{JM-DP} & \multicolumn{3}{c}{FCS-RF} & \multicolumn{1}{c}{Full} & \multicolumn{1}{c}{CCA} \\$\Nbind$ & $\rho$ & $\tau$ & $M$ &  $\bar{U}$ &  $B$ &  $T$ &  $\bar{U}$ &  $B$ &  $T$ & $T$ & $T$ \\ 
  \hline
100 & 0.3 & 0.1 & 1 & 0.03 & 0.00 & 0.03 & 0.03 & 0.00 & 0.03 & 0.02 & 0.04 \\ 
  100 & 0.3 & 0.1 & 50 & 0.03 & 0.05 & 0.07 & 0.03 & 0.06 & 0.09 & 0.02 & 0.04 \\ 
  100 & 0.3 & 0.3 & 1 & 0.05 & 0.00 & 0.05 & 0.05 & 0.00 & 0.05 & 0.02 & 0.09 \\ 
  100 & 0.3 & 0.3 & 50 & 0.05 & 0.19 & 0.24 & 0.05 & 0.18 & 0.23 & 0.02 & 0.09 \\ 
  100 & 0.3 & 0.5 & 1 & 0.11 & 0.00 & 0.11 & 0.08 & 0.00 & 0.08 & 0.02 & 0.15 \\ 
  100 & 0.3 & 0.5 & 50 & 0.11 & 0.33 & 0.43 & 0.08 & 0.31 & 0.39 & 0.02 & 0.15 \\ 
  100 & 0.6 & 0.1 & 1 & 0.03 & 0.00 & 0.03 & 0.03 & 0.00 & 0.03 & 0.03 & 0.04 \\ 
  100 & 0.6 & 0.1 & 50 & 0.03 & 0.05 & 0.08 & 0.03 & 0.06 & 0.09 & 0.03 & 0.04 \\ 
  100 & 0.6 & 0.3 & 1 & 0.05 & 0.00 & 0.05 & 0.05 & 0.00 & 0.05 & 0.03 & 0.10 \\ 
  100 & 0.6 & 0.3 & 50 & 0.05 & 0.19 & 0.24 & 0.05 & 0.17 & 0.21 & 0.03 & 0.10 \\ 
  100 & 0.6 & 0.5 & 1 & 0.10 & 0.00 & 0.10 & 0.08 & 0.00 & 0.08 & 0.03 & 0.15 \\ 
  100 & 0.6 & 0.5 & 50 & 0.10 & 0.33 & 0.43 & 0.08 & 0.30 & 0.37 & 0.03 & 0.15 \\ 
  200 & 0.3 & 0.1 & 1 & 0.01 & 0.00 & 0.01 & 0.01 & 0.00 & 0.01 & 0.01 & 0.02 \\ 
  200 & 0.3 & 0.1 & 50 & 0.01 & 0.04 & 0.05 & 0.01 & 0.05 & 0.07 & 0.01 & 0.02 \\ 
  200 & 0.3 & 0.3 & 1 & 0.01 & 0.00 & 0.01 & 0.02 & 0.00 & 0.02 & 0.01 & 0.05 \\ 
  200 & 0.3 & 0.3 & 50 & 0.01 & 0.15 & 0.16 & 0.02 & 0.17 & 0.19 & 0.01 & 0.05 \\ 
  200 & 0.3 & 0.5 & 1 & 0.03 & 0.00 & 0.03 & 0.03 & 0.00 & 0.03 & 0.01 & 0.11 \\ 
  200 & 0.3 & 0.5 & 50 & 0.04 & 0.29 & 0.33 & 0.04 & 0.30 & 0.33 & 0.01 & 0.11 \\ 
  200 & 0.6 & 0.1 & 1 & 0.02 & 0.00 & 0.02 & 0.02 & 0.00 & 0.02 & 0.02 & 0.02 \\ 
  200 & 0.6 & 0.1 & 50 & 0.02 & 0.04 & 0.06 & 0.02 & 0.05 & 0.07 & 0.02 & 0.02 \\ 
  200 & 0.6 & 0.3 & 1 & 0.02 & 0.00 & 0.02 & 0.02 & 0.00 & 0.02 & 0.02 & 0.06 \\ 
  200 & 0.6 & 0.3 & 50 & 0.02 & 0.14 & 0.16 & 0.02 & 0.16 & 0.18 & 0.02 & 0.06 \\ 
  200 & 0.6 & 0.5 & 1 & 0.03 & 0.00 & 0.03 & 0.03 & 0.00 & 0.03 & 0.02 & 0.12 \\ 
  200 & 0.6 & 0.5 & 50 & 0.03 & 0.28 & 0.31 & 0.03 & 0.28 & 0.32 & 0.02 & 0.12 \\  
   \hline
  \end{tabular}
  \end{subtable}
\end{sidewaystable}

\subsection*{Influence of $\Nbtab$}
\subsubsection*{Accuracy}
\begin{figure}[H]
\begin{center}
\begin{subfigure}[t]{.5\textwidth}
\includegraphics[trim={.4cm .7cm 1.6cm 1cm},scale=.45,clip,height=10cm,width=7cm]{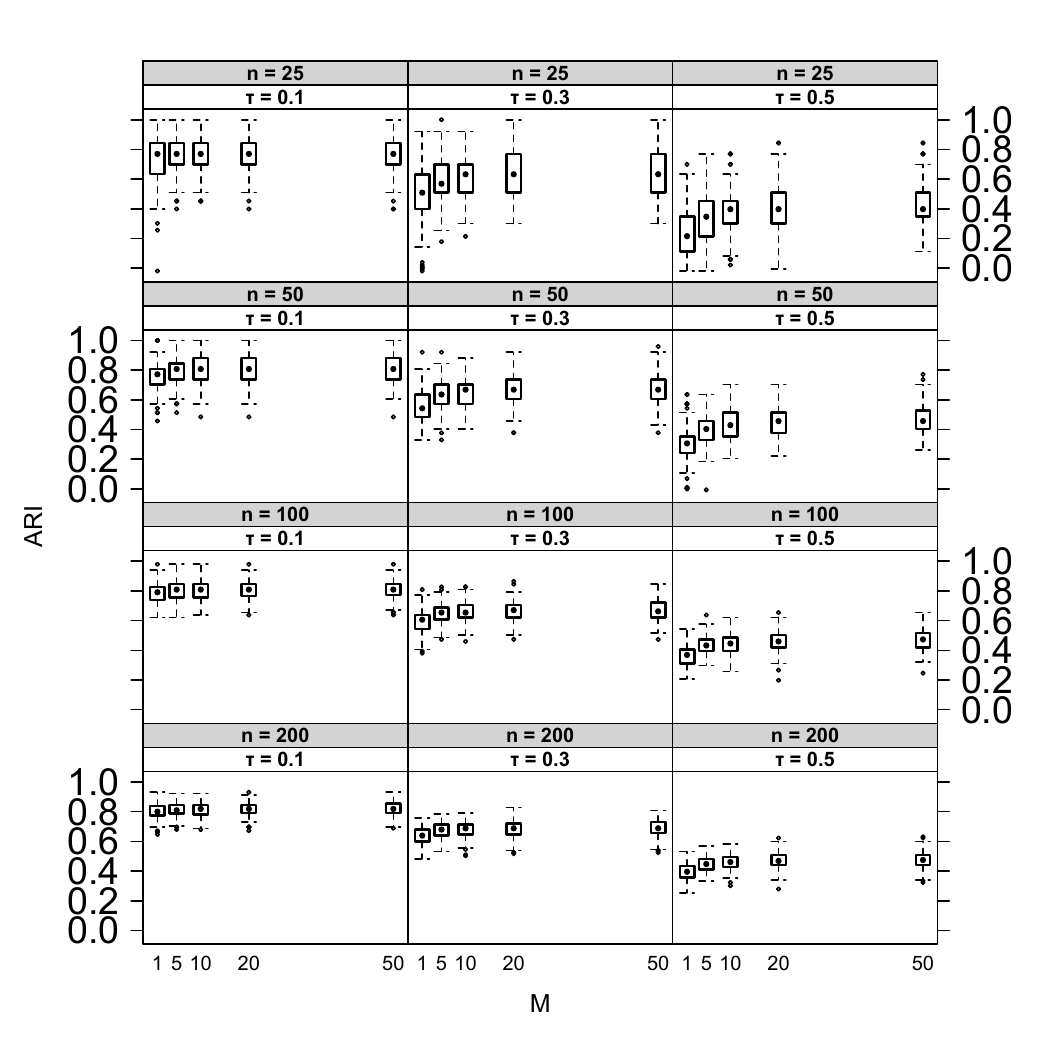}
\caption{ARI distribution for $\rho=0.3$}
\end{subfigure}\begin{subfigure}[t]{.5\textwidth}
\includegraphics[trim={2.2cm .7cm .65cm 1cm},clip,height=10cm,width=7cm]{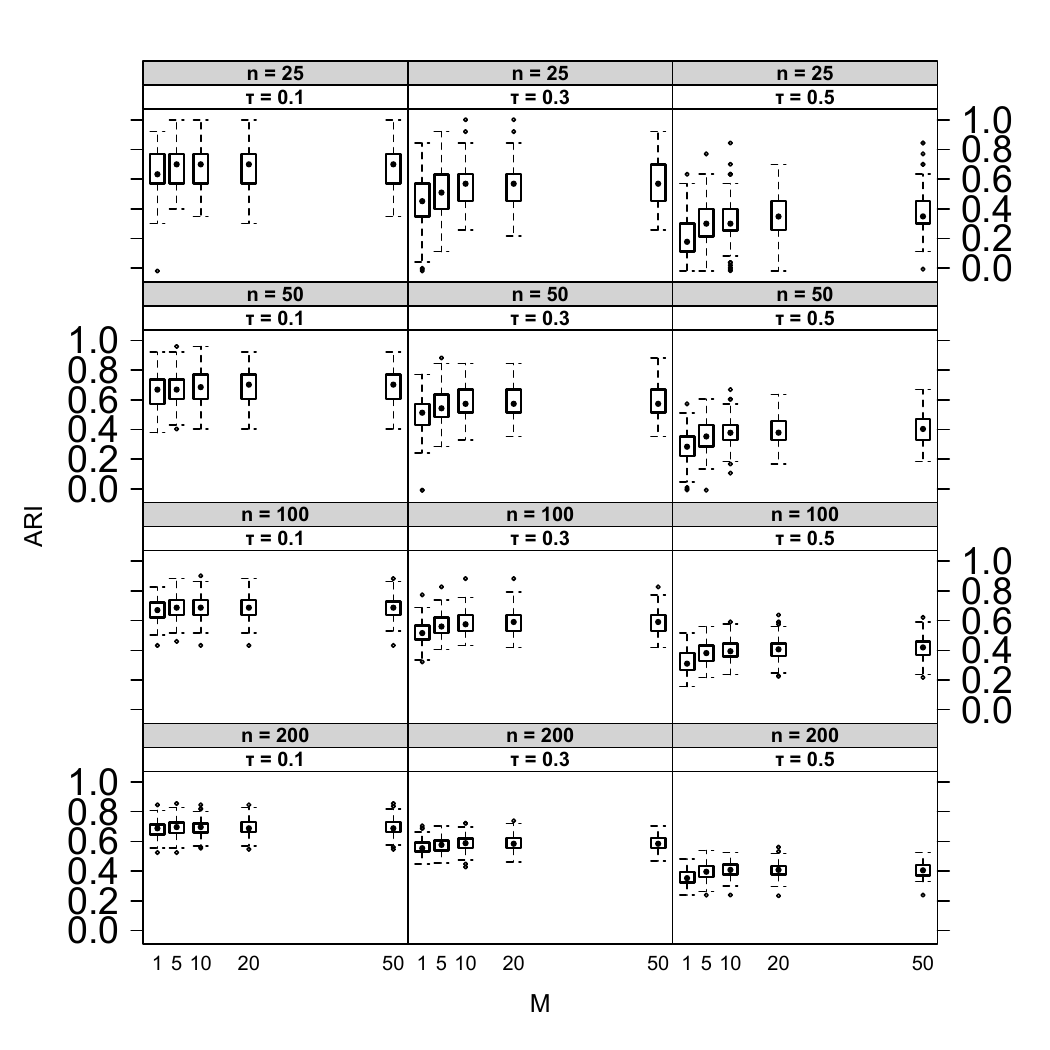}
\caption{ARI distribution for $\rho=0.6$}
\end{subfigure}
\caption{Accuracy of the clustering procedure according to $\Nbtab$: adjusted rand index over the $\Nbsim=200$ generated data sets varying by the number of individuals ($\Nbind$), the correlation between variables ($\rho$) and the proportion of missing values ($\tau$) generated under a MAR mechanism. Data sets are imputed by JM-DP varying by the number of imputed data sets ($\Nbtab$). For each data set, clustering is performed using k-means clustering and consensus clustering is performed using NMF.\label{figM_ari_marmixture}}
\end{center}
\end{figure}

\begin{figure}[H]
\begin{center}
\begin{subfigure}[t]{.5\textwidth}
\includegraphics[trim={.4cm .7cm 1.6cm 1cm},scale=.45,clip,height=10cm,width=7cm]{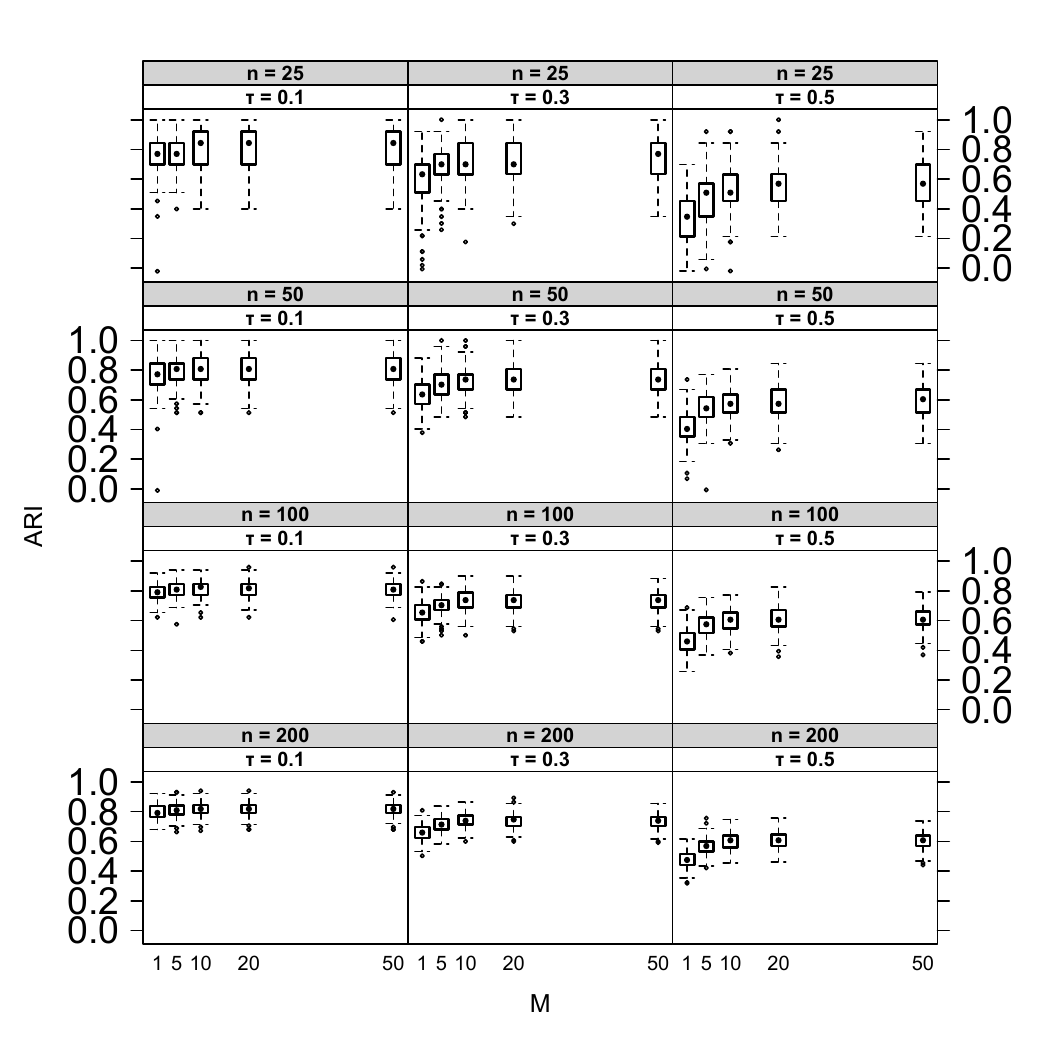}
\caption{ARI distribution for $\rho=0.3$}
\end{subfigure}\begin{subfigure}[t]{.5\textwidth}
\includegraphics[trim={2.2cm .7cm .65cm 1cm},clip,height=10cm,width=7cm]{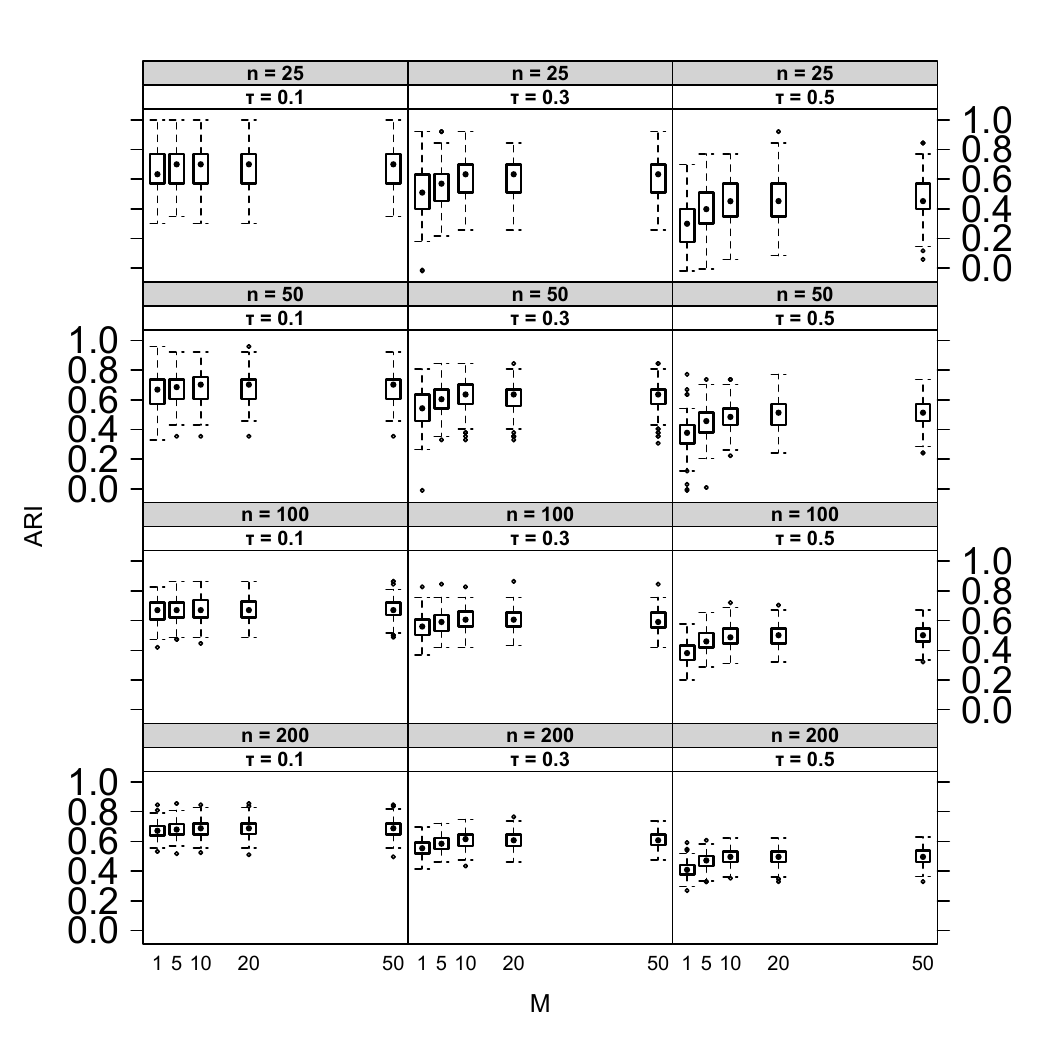}
\caption{ARI distribution for $\rho=0.6$}
\end{subfigure}
\caption{Accuracy of the clustering procedure according to $\Nbtab$: adjusted rand index over the $\Nbsim=200$ generated data sets varying by the number of individuals ($\Nbind$), the correlation between variables ($\rho$) and the proportion of missing values ($\tau$) generated under a MCAR mechanism. Data sets are imputed by FCS-RF varying by the number of imputed data sets ($\Nbtab$). For each data set, clustering is performed using k-means clustering and consensus clustering is performed using NMF.\label{figM_ari_mcarrf}}
\end{center}
\end{figure}

\begin{figure}[H]
\begin{center}
\begin{subfigure}[t]{.5\textwidth}
\includegraphics[trim={.4cm .7cm 1.6cm 1cm},scale=.45,clip,height=10cm,width=7cm]{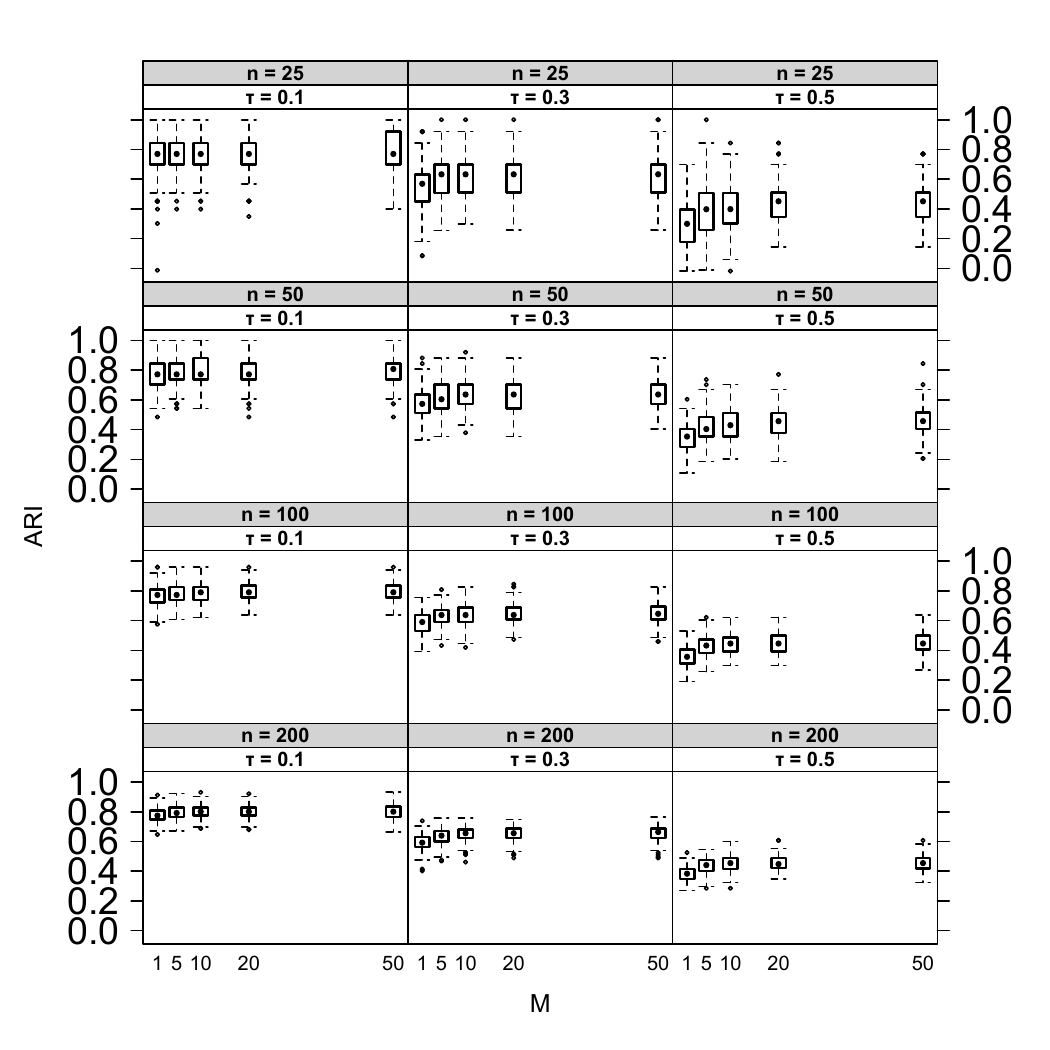}
\caption{ARI distribution for $\rho=0.3$}
\end{subfigure}\begin{subfigure}[t]{.5\textwidth}
\includegraphics[trim={2.2cm .7cm .65cm 1cm},clip,height=10cm,width=7cm]{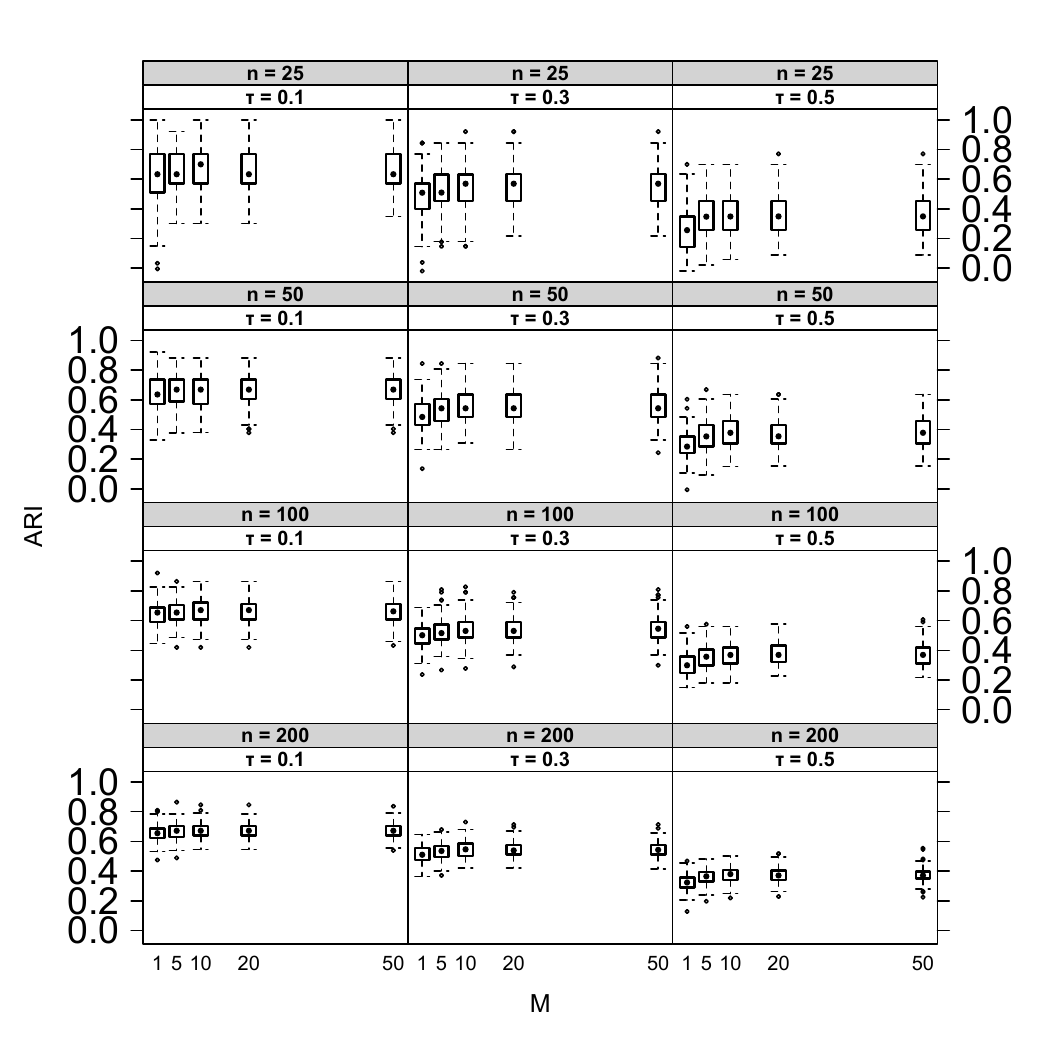}
\caption{ARI distribution for $\rho=0.6$}
\end{subfigure}
\caption{Accuracy of the clustering procedure according to $\Nbtab$: adjusted rand index over the $\Nbsim=200$ generated data sets varying by the number of individuals ($\Nbind$), the correlation between variables ($\rho$) and the proportion of missing values ($\tau$) generated under a MAR mechanism. Data sets are imputed by FCS-RF varying by the number of imputed data sets ($\Nbtab$). For each data set, clustering is performed using k-means clustering and consensus clustering is performed using NMF.\label{figM_ari_marrf}}
\end{center}
\end{figure}
\subsubsection*{Instability}

\begin{figure}[H]
\begin{center}
\begin{subfigure}[t]{.5\textwidth}
\includegraphics[trim={.4cm .7cm 1.6cm 1cm},scale=.45,clip,height=10cm,width=7cm]{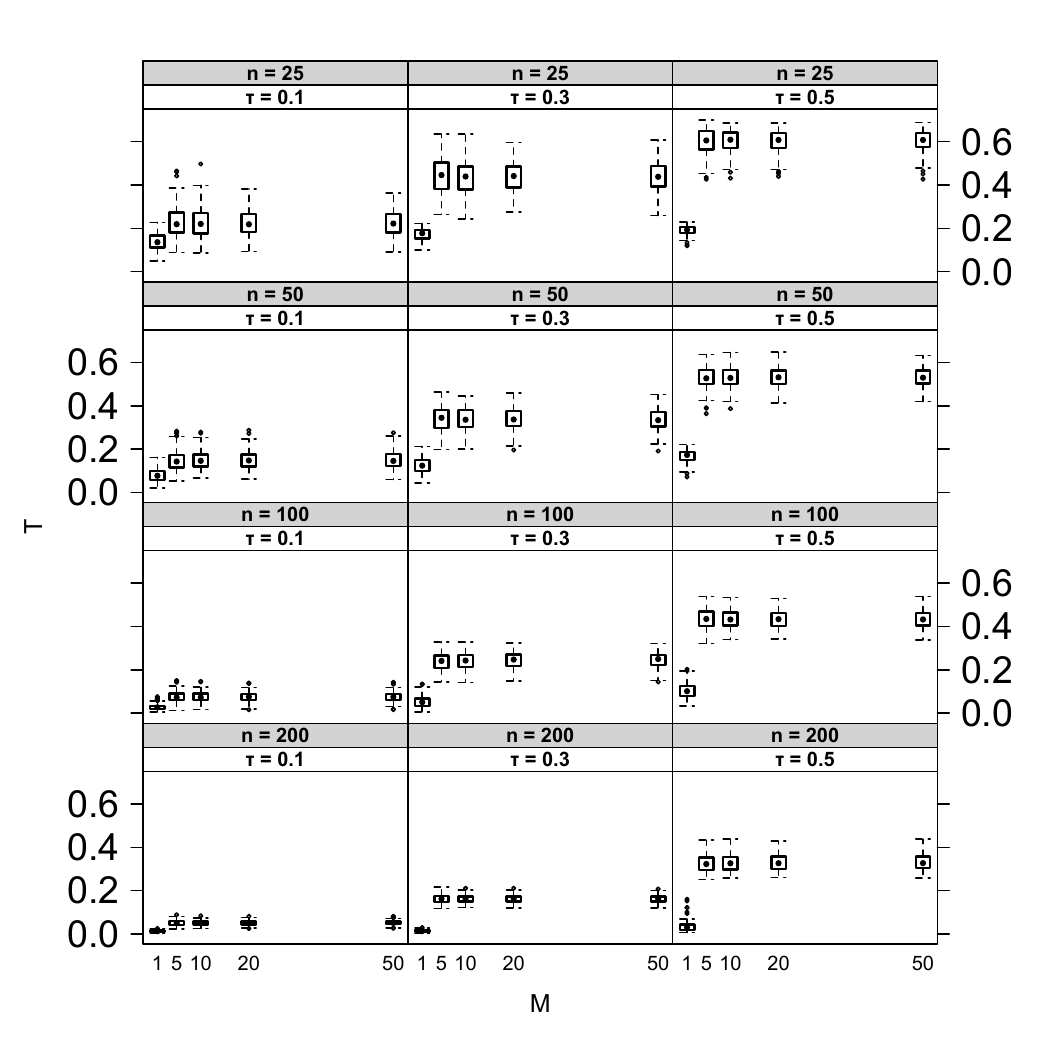}
\caption{$T$ distribution for $\rho=0.3$}
\end{subfigure}\begin{subfigure}[t]{.5\textwidth}
\includegraphics[trim={2.2cm .7cm .65cm 1cm},clip,height=10cm,width=7cm]{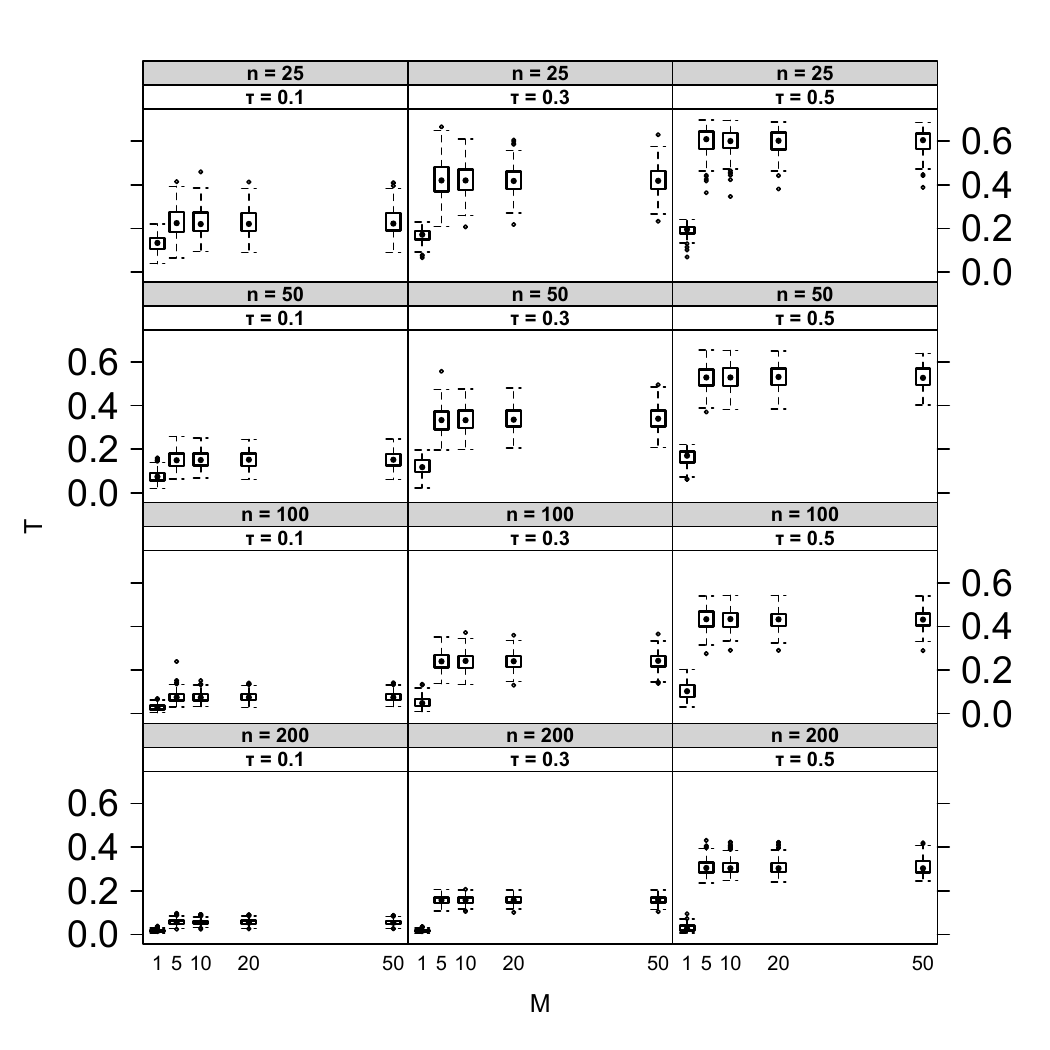}
\caption{$T$ distribution for $\rho=0.6$}
\end{subfigure}
\caption{Instability according to $\Nbtab$: total instability ($T$) over the $\Nbsim=200$ generated data sets varying by the number of individuals ($\Nbind$), the correlation between variables ($\rho$) and the proportion of missing values ($\tau$) generated under a MAR mechanism. Data sets are imputed by JM-DP varying by the number of imputed data sets ($\Nbtab$). For each data set, clustering is performed using k-means clustering and consensus clustering is performed using NMF.\label{figM_T_mixturemar}}
\end{center}
\end{figure}

\begin{figure}[H]
\begin{center}
\begin{subfigure}[t]{.5\textwidth}
\includegraphics[trim={.4cm .7cm 1.6cm 1cm},scale=.45,clip,height=10cm,width=7cm]{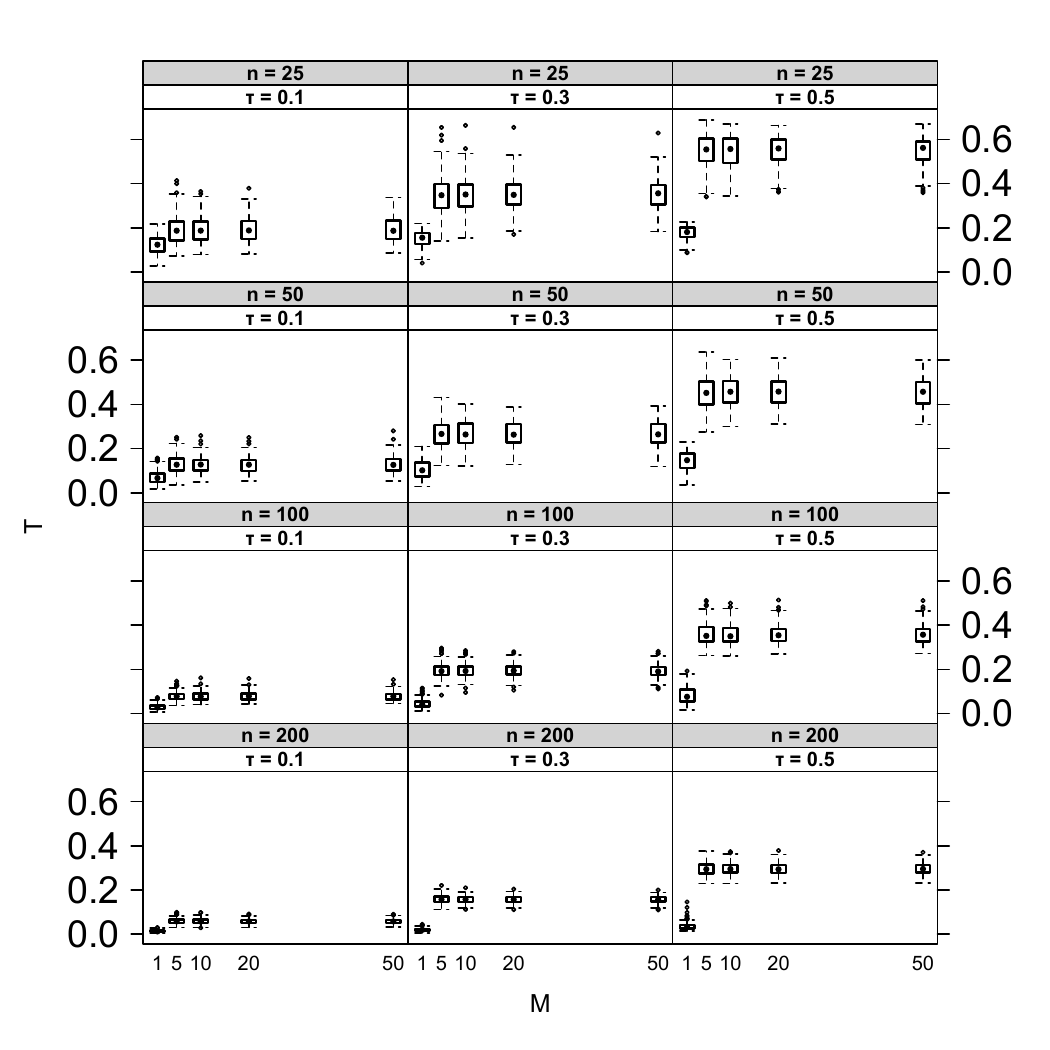}
\caption{$T$ distribution for $\rho=0.3$}
\end{subfigure}\begin{subfigure}[t]{.5\textwidth}
\includegraphics[trim={2.2cm .7cm .65cm 1cm},clip,height=10cm,width=7cm]{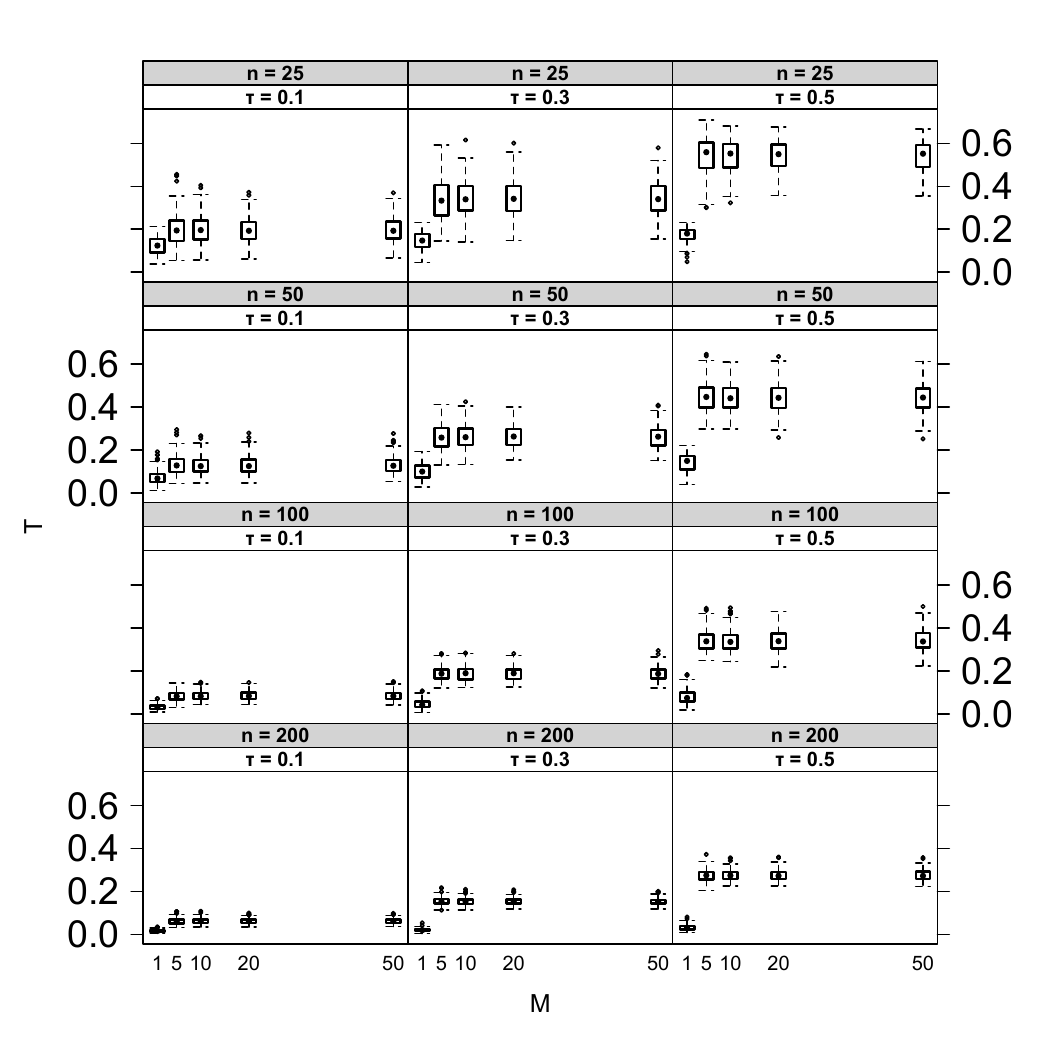}
\caption{$T$ distribution for $\rho=0.6$}
\end{subfigure}
\caption{Instability according to $\Nbtab$: total instability ($T$) over the $\Nbsim=200$ generated data sets varying by the number of individuals ($\Nbind$), the correlation between variables ($\rho$) and the proportion of missing values ($\tau$) generated under a MCAR mechanism. Data sets are imputed by FCS-RF varying by the number of imputed data sets ($\Nbtab$). For each data set, clustering is performed using k-means clustering and consensus clustering is performed using NMF.\label{figM_T_rfmcar}}
\end{center}
\end{figure}

\begin{figure}[H]
\begin{center}
\begin{subfigure}[t]{.5\textwidth}
\includegraphics[trim={.4cm .7cm 1.6cm 1cm},scale=.45,clip,height=10cm,width=7cm]{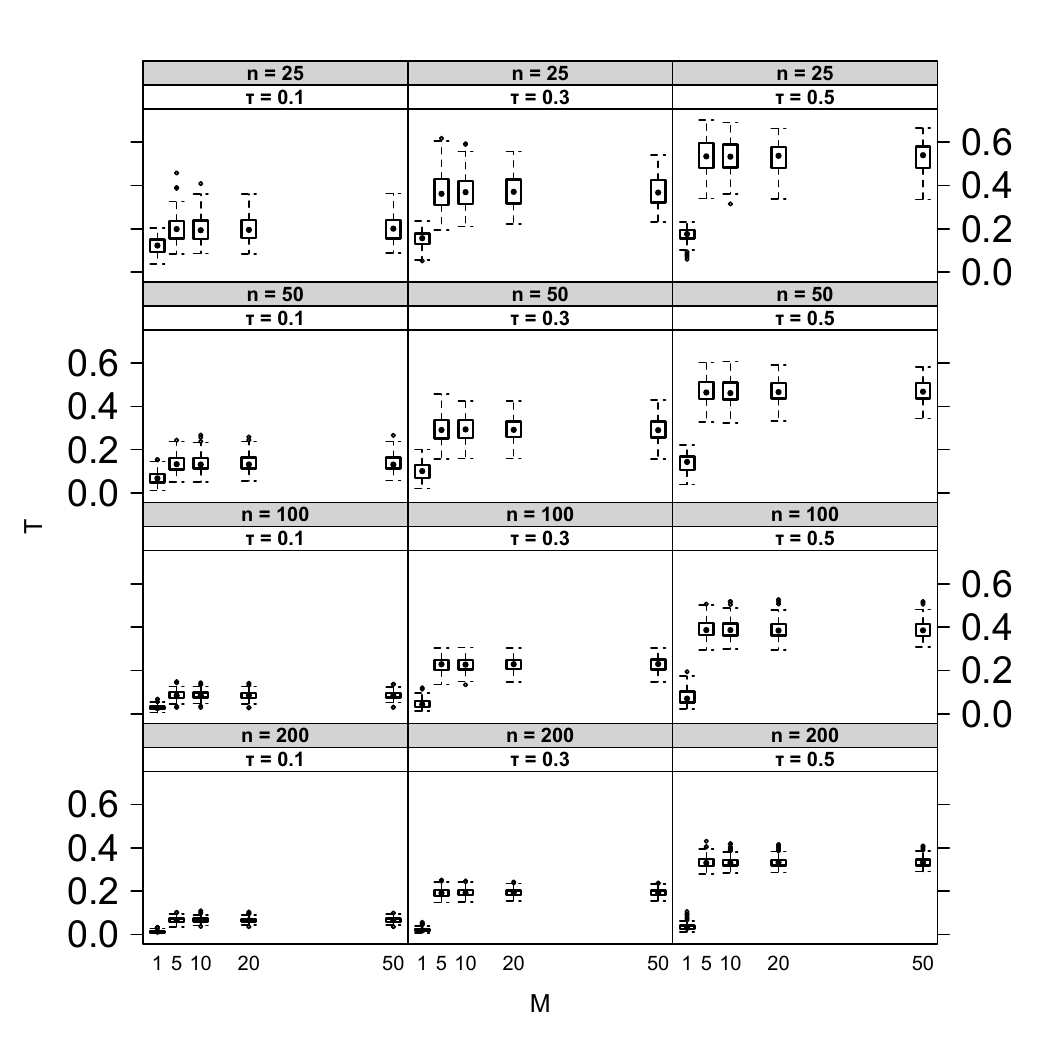}
\caption{$T$ distribution for $\rho=0.3$}
\end{subfigure}\begin{subfigure}[t]{.5\textwidth}
\includegraphics[trim={2.2cm .7cm .65cm 1cm},clip,height=10cm,width=7cm]{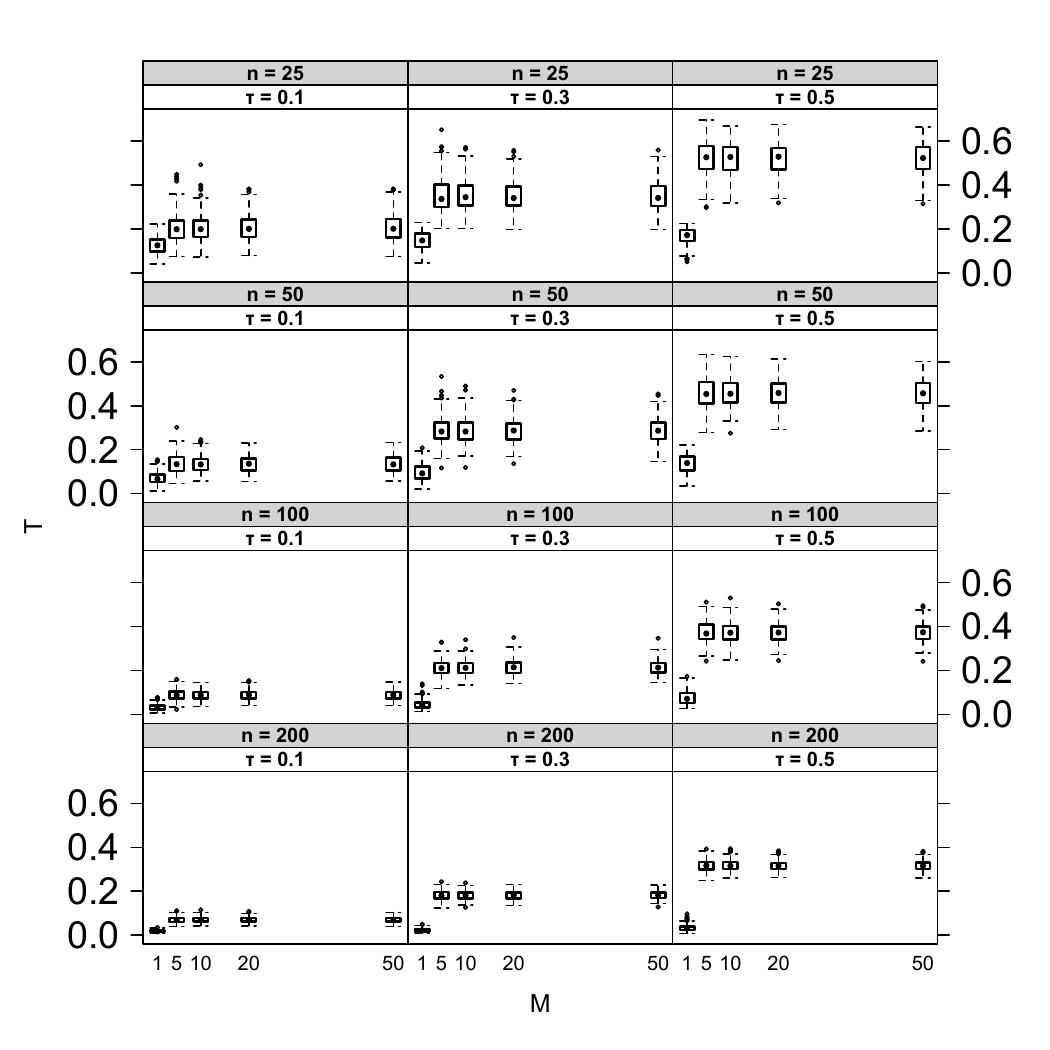}
\caption{$T$ distribution for $\rho=0.6$}
\end{subfigure}
\caption{Instability according to $\Nbtab$: total instability ($T$) over the $\Nbsim=200$ generated data sets varying by the number of individuals ($\Nbind$), the correlation between variables ($\rho$) and the proportion of missing values ($\tau$) generated under a MAR mechanism. Data sets are imputed by FCS-RF varying by the number of imputed data sets ($\Nbtab$). For each data set, clustering is performed using k-means clustering and consensus clustering is performed using NMF.\label{figM_T_rfmar}}
\end{center}
\end{figure}

\subsection*{Application}
\begin{table}[H]
\centering
\caption{Animals data set \label{animals}}
\begin{tabular}{rrrrrrr}
  \hline
 & war & fly & ver & end & gro & hai \\ 
  \hline
ant & 0 & 0 & 0 & 0 & 1 & 0 \\ 
  bee & 0 & 1 & 0 & 0 & 1 & 1 \\ 
  cat & 1 & 0 & 1 & 0 & 0 & 1 \\ 
  cpl & 0 & 0 & 0 & 0 & 0 & 1 \\ 
  chi & 1 & 0 & 1 & 1 & 1 & 1 \\ 
  cow & 1 & 0 & 1 & 0 & 1 & 1 \\ 
  duc & 1 & 1 & 1 & 0 & 1 & 0 \\ 
  eag & 1 & 1 & 1 & 1 & 0 & 0 \\ 
  ele & 1 & 0 & 1 & 1 & 1 & 0 \\ 
  fly & 0 & 1 & 0 & 0 & 0 & 0 \\ 
  fro & 0 & 0 & 1 & 1 &  & 0 \\ 
  her & 0 & 0 & 1 & 0 & 1 & 0 \\ 
  lio & 1 & 0 & 1 &  & 1 & 1 \\ 
  liz & 0 & 0 & 1 & 0 & 0 & 0 \\ 
  lob & 0 & 0 & 0 & 0 &  & 0 \\ 
  man & 1 & 0 & 1 & 1 & 1 & 1 \\ 
  rab & 1 & 0 & 1 & 0 & 1 & 1 \\ 
  sal & 0 & 0 & 1 & 0 &  & 0 \\ 
  spi & 0 & 0 & 0 &  & 0 & 1 \\ 
  wha & 1 & 0 & 1 & 1 & 1 & 0 \\   \hline
\end{tabular}
\end{table}

\end{document}